\documentclass[a4paper,11pt]{article}
\pdfoutput=1 

\usepackage{jheppub} 

\usepackage[T1]{fontenc} 
\usepackage[utf8]{inputenc}

\usepackage{cancel, multirow, caption, subcaption, amsmath}

\graphicspath{{figs/}}

\title{\boldmath Supersymmetric Inflation from the Fifth Dimension}

\author{Kaustubh Deshpande}
\author{and Raman Sundrum}
\affiliation{Maryland Center for Fundamental Physics,\\University of Maryland, College park, MD 20742, USA}

\emailAdd{ksd@umd.edu}
\emailAdd{raman@umd.edu}

\abstract{
We 
develop a supersymmetric bi-axion model of high-scale inflation coupled to supergravity, in which the axionic structure originates from, and is protected by, gauge symmetry in an extra dimension. While local supersymmetry (SUSY)  is necessarily Higgsed at high scales during inflation we show that it can naturally survive down to the $\sim$ TeV scale in the current era in order to resolve the electroweak hierarchy problem. 
We show how a suitable inflationary effective potential for the axions can be generated at tree-level by charged fields under the higher-dimensional gauge symmetry. 
The inflationary trajectory lies along the lightest direction in the bi-axion field space, with  periodic effective potential and 
an effective super-Planckian field range emerging from fundamentally sub-Planckian dynamics. The heavier direction in the field space is shown to also play an important role, as the dominant source of super-Higgsing during inflation.
This model presents an interesting interplay of tuning considerations relating the electroweak hierarchy, cosmological constant and 
inflationary superpotential, where maximal naturalness favors SUSY breaking near the electroweak scale after inflation.
The scalar superpartner of the axionic inflaton, the ``sinflaton'', can naturally have $\sim$ Hubble mass during inflation and sufficiently strong coupling to the inflaton to mediate primordial non-Gaussianities of observable strength in future 21-cm surveys. 
Non-minimal charged fields under the higher-dimensional gauge symmetry can contribute to periodic modulations in the CMB, within the sensitivity of ongoing measurements.
}

\begin{document} 

\preprint{UMD-PP-019-01}

\maketitle
\flushbottom

\section{Introduction}
\label{sec:intro}

Cosmic inflation provides an attractive framework for understanding the robustness of the early state of our
universe (see \cite{Baumann:2009ds} for a review). Its simplest implementation driven by a slowly rolling scalar field (inflaton) requires a very flat inflaton potential, suggesting 
that the inflaton is a 
 pseudo-Goldstone boson of a spontaneously broken global symmetry.  
  A small explicit breaking of the symmetry can then give rise to a weak potential naturally varying on the scale of the spontaneous breaking, $f$. A canonical example is given by the model of ``Natural Inflation''  \cite{PhysRevLett.65.3233}, with 
 periodic inflaton potential,
\begin{equation}
\label{eq:natural_inflation}
V(\phi) = V_0 \left( 1 - \cos \frac{\phi}{f} \right).
\end{equation}
 However, even a crude fit to the Cosmic Microwave Background (CMB) data \cite{Ade:2015lrj} requires 
  $f \gtrsim M_{\rm Pl}$\footnote{This is an example of the model-independent Lyth bound \cite{Lyth:1996im} in the case of Natural Inflation model.}, which conflicts with our general expectation that there should be no dynamical scales above the Planck scale, and with the particular arguments that global symmetries themselves are ill-defined in the context of Quantum Gravity 
\cite{Kallosh:1995hi, Banks:2010zn, Harlow:2018jwu}.

These concerns can be resolved by (a) relating but not identifying 
 the scale over which the inflaton potential varies with the scale of spontaneous symmetry breaking,  
 and (b) realizing the spontaneously broken approximate symmetries as accidental symmetries in the IR rather than fundamental global symmetries in the UV. The simplest version of (a) is given by  beginning with two
pseudo-Goldstone bosons, $\phi_A$ and $\phi_B$, for two global symmetries $U(1)_A \times U(1)_B$ spontaneously broken at approximately the same scale $f_A, f_B \approx f  \ll M_{\rm Pl}$ \cite{Kim:2004rp}. For suitable 
  explicit symmetry breaking sources one can then generate a potential of the form 
\begin{equation}
\label{eq:biaxion_potential}
V(\phi_A, \phi_B) = V_0^{(1)} \left( 1 - \cos \frac{\phi_B}{f_B} \right) + V_0^{(2)} \left[ 1 - \cos \left( \frac{\phi_A}{f_A} + N \frac{\phi_B}{f_B} \right) \right],
\end{equation}
where $N$ represents a large charge under $U(1)_B$ for one of the ``spurions'' characterizing the explicit breaking. Naively, this makes the problem worse, since the potential varies in the $\phi_A$ direction on the scale $f \ll M_{\rm Pl}$, and in the $\phi_B$ direction on the scale $f/N \ll f$, while CMB data suggests a potential varying more slowly than the Planck scale. However, just such a potential can arise when we properly consider the mass eigenstates. Taking for simplicity $V_0^{(1), (2)} \approx V_0$, these are given by heavy and light directions in field space,
\begin{equation}
\phi_h \equiv \phi_B + \frac{1}{N} \phi_A , \ \phi_l \equiv \phi_A - \frac{1}{N} \phi_B. 
\end{equation} 
After setting the heavy $\phi_h$ to its vacuum expectation value (VEV), we can obtain the effective potential for the light field $\phi_l$ as
\begin{equation}
\label{eq:natural_inflation_from_biaxion}
V_{\rm eff}(\phi_l) \big\rvert_{\langle \phi_h \rangle \approx 0} \approx V_0 \left( 1 - \cos \frac{\phi_l}{N f} \right).
\end{equation}
This corresponds to an effective Natural Inflation model, with inflaton  $\phi_l$ and an emergent scale of potential variation 
 $f_{\rm eff} = N f$, which can be $> M_{\rm Pl}$ even though $f < M_{\rm Pl}$, for sufficiently large spurious charge $N$. We will refer to this as the ``Bi-axion inflation'' model. 

An attractive microscopic realization of Bi-axion inflation satisfying (b), based on the mechanism of  ``extranatural inflation'' \cite{ArkaniHamed:2003wu}, is  provided by using gauge symmetry in an extra dimension \cite{Bai:2014coa}. If the higher-dimensional spacetime is highly warped so as to have an AdS$_5$/CFT$_4$ type holographic purely-4D dual description, then the dual interpretation is that the axions are composite Goldstone bosons of some strong dynamics (see e.g. \cite{Contino:2003ve}), analogous to the pions of QCD, and the spontaneously broken symmetries are accidental or emergent symmetries below the Planck scale. Here, we just briefly summarize the unwarped (or mildly warped) higher-dimensional case. 
The 4D axions above are realized as gauge-invariant Wilson-loops (or lines, given suitable boundary conditions) around (or across) the compact extra dimension,
\begin{equation}
\phi_A \equiv \int_0^L A_5 \ dx_5 \ , \ \phi_B \equiv \int_0^L B_5 \ dx_5.
\end{equation}
Charged matter propagating in the 5D bulk, $H_1$ and $H_2$, with mass $m$, can generate the potential \eqref{eq:biaxion_potential} for $\phi_A$ and $\phi_B$, given that they are charged under the two gauge groups as $(0, 1)$ and $(1, N)$, respectively. The scales $f_A, f_B$ emerge as
\begin{equation}
f_{A} = \frac{1}{g_{A} L}, \ f_{B} = \frac{1}{g_{B} L}.
\end{equation}
The potential in \eqref{eq:biaxion_potential} can be generated minimally by the loop contributions of $H_1, H_2$ via the ``Hosotani mechanism'' \cite{HOSOTANI1983193} which gives
\begin{equation}
V_0^{\rm loop} \sim \frac{e^{-mL}}{L^4},
\end{equation}
in \eqref{eq:natural_inflation_from_biaxion}, as well as ``higher harmonics'' accompanied by higher powers of $e^{-mL}$.
As studied in \cite{delaFuente:2014aca},  bi-axion extranatural inflation can also non-trivially satisfy the plausible constraints of the Weak Gravity Conjecture (WGC) \cite{ArkaniHamed:2006dz}. These quantum gravity constraints are an even stronger form of the arguments forbidding fundamental global symmetries, to also forbid UV gauge symmetries with very weak gauge couplings (relative to gravitational strength).
These higher dimensional realizations of bi-axion inflation can be generalized to multiple-axion models, which then allow for more modest  values of charge, $N$ \cite{Bai:2014coa, delaFuente:2014aca}.


In this paper, we study compatibility of the bi-axion inflation scenario arising from higher dimensional gauge theory with the scenario of $\sim$TeV-scale supersymmetry (SUSY) for resolving the electroweak hierarchy problem. 
In the presence of SUSY, the loop contributions from the charged matter fields to the effective potential of 4D axions cancel out. We are hence forced to have tree-level contributions for the same, which can be achieved if $H_1, H_2$ have non-zero VEVs ($v, v^\prime$) at both the boundaries, which generates 
\begin{equation}
V_0^{\rm tree} \sim e^{-mL} m v v^\prime.
\end{equation}
Obviously, the question of whether the above-mentioned very plausible and robust forms of inflation are naturally realizable within the constraints of supergravity (SUGRA) dynamics in the UV, with SUSY being present at $\sim$ collider energies today, is of considerable importance to our picture of fundamental physics and the prospects for experiments and observations. 
See \cite{Czerny:2014qqa, Gao:2014uha, Long:2014dta, Ali:2014mra, Ben-Dayan:2014lca, Palti:2015xra, Kappl:2015pxa} for other discussions of bi-axion inflation combined with SUSY, where the axions have alternative UV realizations. 
See \cite{Kallosh:2004yh, He:2010uk, Kobayashi:2010rx, Antusch:2011wu, Yamada:2012tj, Czerny:2014xja, Buchmuller:2015oma} for other attempts to reconcile low energy SUSY and inflation from a UV perspective.
We will also explore the possible new signatures from extra fields in the axion supermultiplets, most notably in the form of primordial non-Gaussianities (NG) in the cosmological collider physics program \cite{Chen:2009we, Chen:2009zp, Chen:2010xka, Baumann:2011nk, Assassi:2012zq, Chen:2012ge, Pi:2012gf, Noumi:2012vr, Arkani-Hamed:2015bza, Dimastrogiovanni:2015pla, Kumar:2017ecc} as well as periodic modulations in the CMB \cite{Wang:2002hf, Pahud:2008ae, Flauger:2009ab, Kobayashi:2010pz, Easther:2013kla, Flauger:2014ana, delaFuente:2014aca, Higaki:2014mwa, Choi:2015aem, Price:2015xwa}. 

Models of single-field inflation with relatively simple potentials, such as Natural Inflation and its variants, necessarily operate at 
high scales  in order to satisfy cosmological data, with inflationary Hubble scale $H_{\rm inf} \sim 10^{13} - 10^{14}$ GeV. 
The recently released Planck 2018 data places tight constraints on such high-scale models, especially given 
the non-observation of CMB B-modes induced by super-horizon gravitational waves  \cite{Akrami:2018odb}. Natural Inflation itself 
 is now disfavored  at 95\% confidence level, but not ruled out. However, 
 the bi-axionic structure of inflation from extra-dimensional gauge symmetry can generically produce multiple periodic terms in the potential \eqref{eq:natural_inflation}, which  can alleviate the tension above with a suitable and plausibly not very fine-tuned choice of parameters. We leave such a detailed analysis and appraisal for a future study. Furthermore, there are various ways discussed in the existing literature to relax these constraints for axion-based inflation,  e.g. by realizing the structure of hybrid inflation from a bi-axion potential \cite{Peloso:2015dsa}.	

The paper is organized as follows. In Section~\ref{sec:KL_toy_model}, we review a SUGRA-based inflation model, the ``Kallosh-Linde-Rube model'' \cite{Kawasaki:2000yn, Kallosh:2010xz}, which has many common features with our SUSY bi-axion model as developed in Sections~\ref{sec:sugra_biaxion} and \ref{sec:inflationary_history}. In Section~\ref{sec:sugra_biaxion}, starting from the 5D SUSY gauge structure, we first construct a 4D effective theory of an axion supermultiplet. After generalizing it to the case of two axions, we account for (effective) 4D SUGRA couplings below the compactification scale. 
In Section~\ref{sec:inflationary_history}, we discuss the inflationary trajectory along the lightest direction in the field space with an effective super-Planckian field range and periodic potential, also stabilized along all the other heavier directions. We then describe the picture of SUSY breaking ($\cancel{\textrm{SUSY}}$) during inflation which we find to be caused mostly by the heavy sector and not the inflaton sector. 
Furthermore, we also account for the post-inflationary $\cancel{\textrm{SUSY}}$ vacuum that we occupy today, which we find not affecting the inflationary dynamics significantly as long as the $\cancel{\textrm{SUSY}}$ scale is much below the inflationary energy scale.
This model presents an interesting interplay of fine-tunings in the electroweak (EW) sector, cosmological constant (CC), and superpotential which are connected together after incorporating the $\cancel{\textrm{SUSY}}$ today. The superpotential fine-tuning favors $\cancel{\textrm{SUSY}}$ at high-scale, however the net fine-tuning, dominated by the EW and CC fine-tunings, can be shown to favor $\cancel{\textrm{SUSY}}$ at low-scale i.e. somewhat above the EW scale.
In Section~\ref{sec:observable_signals}, we discuss observable signals in the form of primordial NG and periodic modulations in the CMB. The ``sinflaton'', the real scalar partner of inflaton, can have $\mathcal{O}(H_{\rm inf})$ mass during inflation and sufficiently strong coupling to the inflaton to mediate primordial NG of observable strength in future experiments. A boundary-localized gauge singlet, in the presence of a shift-symmetric K$\ddot{\mathrm{a}}$hler coupling, can also mediate sizeable primordial NG. Charged matter much heavier than the compactification scale, even only modestly below the 5D gauge theory cut-off, can contribute to periodic modulations in the CMB, within the sensitivity of ongoing searches. We conclude in Section~\ref{sec:conclusions}.

We use units with the reduced Planck mass $M_{\rm Pl} = 1$  everywhere in the paper, except Sections~\ref{subsec:finetunings} and \ref{sec:observable_signals}, where we explicitly write factors of $M_{\rm Pl}$ in order to get a better sense of the numbers.

\section{The Kallosh-Linde-Rube model}
\label{sec:KL_toy_model}

We seek a locally supersymmetric description of high-scale inflation in which SUSY is only broken somewhat above the weak scale today. 
Since the weak scale is $\ll H_{\rm inf}$, we can first consider the supersymmetric limit of the ground state today. 
On the other hand, during inflation we know that the approximate de Sitter geometry is incompatible with SUSY. So 
inflation must be a spontaneous breaking (super-Higgsing) of SUSY within an excitation on top of today's SUSY vacuum, which we can also approximate to have zero vacuum energy (cosmological constant).
 
In order to have a light inflaton $(\phi)$, we will have an inflaton supermultiplet $(\Phi)$ with approximate shift symmetry. This can be implemented with $K (\Phi, \bar{\Phi}) = K(\Phi + \bar{\Phi})$ and $\phi = \textrm{Im}(\Phi)$. A small explicit breaking of the shift symmetry from the superpotential can generate slow-roll potential for $\phi$. Thus, the lightness of inflaton can be explained by its pseudo-Goldstone boson nature.
However, implementing inflation with only this single supermultiplet is challenging \cite{Achucarro:2012hg}. In this case, the Goldstino of spontaneous $\cancel{\textrm{SUSY}}$ during inflation would have to be the inflatino (then ``eaten'' by the gravitino). Consider $K = \frac{1}{2} \left(\Phi + \bar{\Phi}\right)^2$ and $W = f(\Phi)$. Then, restricting to polynomial $f(\Phi)$ for illustration,  in SUGRA, $V(\phi) \approx f^{\prime 2} (\phi / \sqrt{2}) - 3 f^2(\phi / \sqrt{2})$, which has a clear instability.\footnote{However, see \cite{Ferrara:2016vzg, Ketov:2014hya, AlvarezGaume:2011xv} for attempts towards building ``sGoldstino inflation'' model.} 
This can be avoided by introducing a separate supermultiplet for the Goldstino during inflation.

We see that the Goldstino multiplet must be part of a sector that Higgses SUSY {\it during} inflation. One of the simplest models to describe spontaneous $\cancel{\textrm{SUSY}}$ coupled to SUGRA, is the Polonyi model \cite{Polonyi:1977pj}:
\begin{equation}
\label{eq:Polonyi_model}
K = \bar{S} S - \lambda \left(\bar{S} S\right)^2 , \ W = \mu S,
\end{equation}
with the addition of the non-minimal K$\ddot{\mathrm{a}}$hler coupling $\lambda$.
The SUSY order parameter in the vacuum is $D_S W\big\rvert_{\langle S \rangle \approx 0} \approx \mu \neq 0$. 
Spontaneous $\cancel{\textrm{SUSY}}$ in this model gives rise to a massless Goldstino which however is eaten by the gravitino which then becomes massive (``super-Higgs mechanism''). The quartic term in the K$\ddot{\mathrm{a}}$hler potential also makes the scalar heavy, with $m_S^2 \approx 4 \lambda \mu^2$. Thus, there is no light particle in this sector. 
During inflation, in the limit of the slow-roll approximation i.e. for a fixed value of inflaton, the physics can be approximately described by this model. 
But we need to have a coupling between this sector $(S)$ and the inflaton  $(\Phi)$ such that there is no $\cancel{\textrm{SUSY}}$ at $\Phi = 0$ (i.e. at the vacuum today) but with $\cancel{\textrm{SUSY}}$ at $\Phi = \Phi_0 \neq 0$ (i.e. during inflation). In other words, the $\mu$ parameter of \eqref{eq:Polonyi_model} needs to be made $\Phi$-dependent in a suitable manner. This can be achieved with the following model \cite{Kawasaki:2000yn, Kallosh:2010xz}:
\begin{equation}
\label{eq:Kallosh_Linde_model_Lagrangian}
K = \frac{1}{2} \left(\Phi + \bar{\Phi}\right)^2 + \bar{S} S - \lambda \left(\bar{S} S\right)^2 , \ W = S f(\Phi),
\end{equation}
which we will refer to as the ``Kallosh-Linde-Rube (KLR) model'' and consider as a toy model for our SUSY bi-axion model.
All scalars except for $\phi = \textrm{Im}(\Phi)$ can be shown to be heavy and thus the inflationary potential (see Appendix~\ref{sec:appendix_sugra_basics} for SUGRA scalar potential) can be obtained as,
\begin{equation}
\begin{gathered}
V(\phi) \big\rvert_{\langle S\rangle , \langle \eta \rangle} = f^2(\phi / \sqrt{2}), \\
V_{\rm inf} = 3 H_{\rm inf}^2 = f^2(\phi_0 / \sqrt{2}).
\end{gathered}
\end{equation}
The SUSY order parameters for $\Phi$ and $S$ during inflation can be evaluated as follows:
\begin{equation}
\label{eq:SUSY_order_param_KL_model}
D_\Phi W \big \rvert_{\rm inf} \approx 0 \ , \ D_S W \big \rvert_{\rm inf} \approx f(\Phi_0) \neq 0.
\end{equation}
This implies that, as expected, $\cancel{\textrm{SUSY}}$ during inflation is caused by the heavy sector $(S)$. Hence the Goldstino during inflation (further eaten by the gravitino) is equal to the fermion from the $S$-sector $(\psi_S)$ and not the inflatino $(\psi_\Phi)$.

The real scalar partner of the inflaton i.e. sinflaton $(\eta = \textrm{Re}(\Phi))$, has the following mass coming from its coupling to the SUSY-breaking curvature from \eqref{eq:Kallosh_Linde_model_Lagrangian}:
\begin{equation}
\label{eq:KL_model_sinflaton_mass}
m_\eta \approx \sqrt{6} H_{\rm inf}.
\end{equation}
This is within the favorable range of masses for observing it in primordial NG in the cosmological collider physics program. 
However, such a light sinflaton (i.e. $m_{\eta} \sim {\mathcal O}(H_{\rm inf})$) is not guaranteed from this class of models. Indeed, a higher order term in the K$\ddot{\mathrm{a}}$hler potential with a direct coupling between $S$ and $\Phi$, respecting the shift symmetry of $\phi$,  
\begin{equation}
K \ni - \frac{c}{\Lambda^2} \left(\Phi + \bar{\Phi}\right)^2 \bar{S} S,
\end{equation}
can give a large contribution to the sinflaton mass:
\begin{equation}
\label{eq:KL_model_sinflaton_mass_additional}
m_\eta^2 \approx 2 V_{\rm inf} + c \frac{V_{\rm inf}}{\Lambda^2} \approx 6 H_{\rm inf}^2 \left( 1 + \frac{c}{2 \Lambda^2} \right).
\end{equation}
Thus, $m_{\eta} \sim {\mathcal O}(H_{\rm inf})$ for $\Lambda \approx \mathcal{O}(1) M_{\rm Pl}$. But, $m_\eta \gg H_{\rm inf}$ is also possible with $\Lambda \ll M_{\rm Pl}$.

Even assuming $m_{\eta} \sim {\mathcal O}(H_{\rm inf})$,  in order for $\eta$ to mediate observable primordial NG, there has to be sufficiently strong coupling between it and the inflaton $(\phi)$. The SUGRA scalar potential from \eqref{eq:Kallosh_Linde_model_Lagrangian} does have such couplings, but these are shift-symmetry violating and hence very small, e.g. $\mathcal{L} \ni m_\phi^2 \eta^2 \phi^2 \sim 10^{-10} \eta^2 \phi^2$. However, higher order shift-symmetric terms in K$\ddot{\mathrm{a}}$hler potential,
\begin{equation}
K \ni \frac{c^\prime}{\Lambda^2} \left( \Phi + \bar{\Phi} \right)^4,
\end{equation}
can generate derivative-interactions as
\begin{equation}
\mathcal{L} \ni \frac{c^\prime}{\Lambda^2} (\partial \phi)^2 \eta^2.
\end{equation}
This sinflaton-inflaton interaction (with a non-zero VEV for $\eta$) along with $m_{\eta} \sim {\mathcal O}(H_{\rm inf})$ can give rise to observable NG for sufficiently small $\Lambda$ and large $\langle\eta\rangle$. However, in this paper, we will not pursue the phenomenology of this model further. 

The main drawback of this construction is that the origin of such a form of Lagrangian \eqref{eq:Kallosh_Linde_model_Lagrangian} is not explained within the model. Also, it suffers from the issue of trans-Planckian field displacement needed for $\phi$, since a typical choice of $f(\Phi)$ in \eqref{eq:Kallosh_Linde_model_Lagrangian} gives a large-field inflation model subject to the Lyth bound \cite{Lyth:1996im}. 

\section{SUSY bi-axion model}
\label{sec:sugra_biaxion}

In this section, we develop the setup of supersymmetric inflation with the pseudo-Goldstone boson (or axion) nature of inflaton derived from a gauge symmetry in a compact extra dimension (``extranatural inflation'' \cite{ArkaniHamed:2003wu}). Firstly, we describe how we obtain the effective theory of a light axion supermultiplet starting from the $\mathcal{N} = 1$ 5D SUSY gauge theory. Then, we describe how to introduce two such axion supermultiplets in order to get $f_{\rm eff} > M_{\rm Pl}$ (for trans-Planckian field displacement satisfying the WGC). Finally, we also discuss how to take into account gravity, thus constructing our ``SUSY bi-axion model''.

As we will see later, this model has many common features with the KLR model described in Section~\ref{sec:KL_toy_model}. It however provides a more UV-complete and robust picture of inflationary dynamics where the central features are governed by the 5D SUSY gauge theory structure.

\subsection{Light axion supermultiplet from 5D SUSY gauge theory}
\label{subsec:5DSUSY_gaugetheory}

\begin{figure}[t]
\centering
\begin{subfigure}{.5\textwidth}
  \centering
  \includegraphics[width=0.9\textwidth]{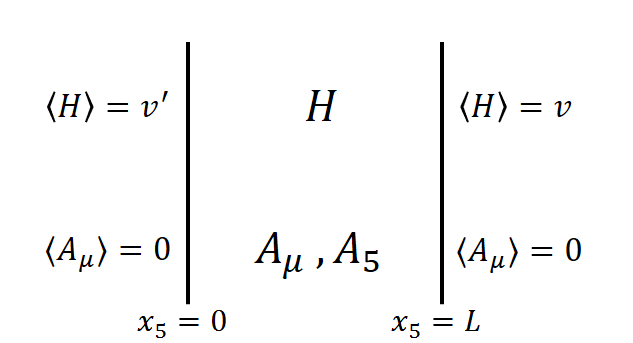}
  \caption{Non-SUSY}
  \label{fig:single_axion_setup_nonSUSY}
\end{subfigure}%
\begin{subfigure}{.5\textwidth}
  \centering
  \includegraphics[width=0.9\textwidth]{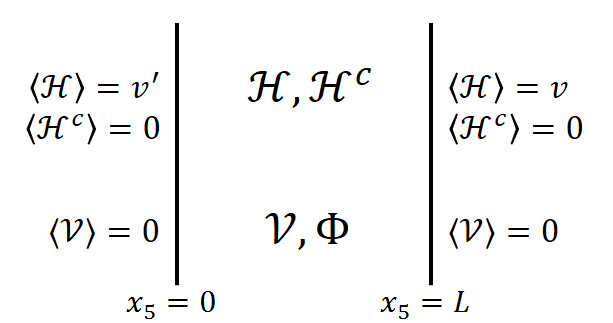}
  \caption{SUSY}
  \label{fig:single_axion_setup_SUSY}
\end{subfigure}
\caption{5D gauge field and charged matter: (a) non-SUSY and (b) SUSY version. See text and Table~\ref{table:5D_SUSY_in_terms_of_4D_superfields} for details.}
\label{fig:single_axion_setup_nonSUSY_SUSY}
\end{figure}

In this sub-section, we will show how a single light axion supermultiplet can emerge from 5D SUSY gauge theory. The extension to the more realistic case of two axion supermultiplets follows in the next sub-section.
Consider a flat extra dimension with boundaries, with a gauge field $A_M$ and a charged scalar field $H$ propagating in the bulk (see Figure~\ref{fig:single_axion_setup_nonSUSY}). If $A_\mu$ and $A_5$ have, respectively, Dirichlet and Neumann boundary conditions at both the boundaries, then only $A_5$ has a zero-mode $\left(A_5^{(0)}\right)$. As mentioned in Section~\ref{sec:intro}, if $H$ has non-zero VEVs at both the boundaries, then it gives a tree-level contribution to the effective potential of $A_5^{(0)}$.

Now, consider the full 5D supersymmetric version of this setup (see Figure~\ref{fig:single_axion_setup_SUSY}). $\mathcal{N} = 1$ 5D SUSY is equivalent to $\mathcal{N} = 2$ 4D SUSY which can be written in the $\mathcal{N} = 1$ 4D SUSY language as follows \cite{ArkaniHamed:2001tb} (see Table~\ref{table:5D_SUSY_in_terms_of_4D_superfields}). 
\begin{table}
\centering
\begin{tabular}{|c|c|c|}
\hline
5D super-multiplet & 5D fields & $\mathcal{N} = 1$ 4D superfields \\
\hline
\multirow{2}{*}{Gauge multiplet} & \multirow{2}{*}{$A_M, \chi_{Dirac}, \eta_{real}$} & Vector superfields: $\mathcal{V}(x_5) \ni A_\mu (x_5), \chi_1 (x_5)$ \\
& & Chiral superfields: $\Phi (x_5) \ni \eta (x_5) + i A_5 (x_5) , \chi_2 (x_5)$ \\
\hline
\multirow{2}{*}{Hypermultiplet} & \multirow{2}{*}{$H, H^c, \psi_{Dirac}$} & Chiral superfields: $\mathcal{H} (x_5) \ni H (x_5), \psi (x_5)$ \\
& & $\mathcal{H}^c (x_5) \ni H^c (x_5) , \psi^c (x_5)$ \\
\hline
\end{tabular}
\caption{$\mathcal{N} = 1$ 5D SUSY in the $\mathcal{N} = 1$ 4D SUSY language}
\label{table:5D_SUSY_in_terms_of_4D_superfields}
\end{table}
5D SUSY gauge multiplet has a gauge field $(A_M)$, Dirac gaugino $(\chi_{Dirac})$ and a real scalar $(\eta)$. These can be represented in $\mathcal{N} = 1$ 4D SUSY language in terms of  vector superfields $\mathcal{V}(x_5) \ni A_\mu (x_5) , \chi_1 (x_5)$ and chiral superfields $\Phi (x_5) \ni \eta (x_5) + i A_5 (x_5) , \chi_2 (x_5)$, where the extra-dimensional coordinate $x_5$ is viewed as a mere continuous ``label'' from the  $\mathcal{N} = 1$ 4D viewpoint.  
 Charged matter fields in 5D SUSY are part of a hypermultiplet which includes two complex scalars which are conjugates of each other under the respective gauge group $(H, H^c)$ and a Dirac fermion $(\psi_{Dirac})$. These can be represented in $\mathcal{N} = 1$ 4D SUSY language in terms of chiral superfields with conjugate representations: $\mathcal{H} (x_5) \ni H (x_5), \psi (x_5) ; \mathcal{H}^c (x_5) \ni H^c (x_5) , \psi^c (x_5)$, again with the continuous ``label'' $x_5$.  

As illustrated in \cite{ArkaniHamed:2001tb}, imposing 4D SUSY and 5D Poincare symmetry automatically generates an emergent 5D SUSY. Thus, the full 5D Lorentz-invariant, gauge-invariant and SUSY action for a gauge multiplet and a charged hypermultiplet, keeping manifest only the $\mathcal{N} = 1$ 4D SUSY, can be written as follows:
\begin{equation}
\label{eq:singleaxion_SUGRA_S5}
\begin{split}
\mathcal{S}_5 =&  \int d^4x \int_0^L dx_5 \left[ \int d^2 \theta \ \frac{1}{4} \mathcal{W}_\alpha^2 + \rm{h.c.} + \int d^4 \theta \ \left\{ \partial_5 \mathcal{V} - \frac{1}{\sqrt{2}} (\Phi + \bar{\Phi}) \right\}^2 \right.\\
+& \left. \int d^4 \theta \ \left( \mathcal{H}^c e^{g_5 \mathcal{V}} \bar{\mathcal{H}}^c + \bar{\mathcal{H}} e^{-g_5 \mathcal{V}} \mathcal{H} \right) + \int d^2 \theta \ \left\{ \mathcal{H}^c \left(m + \partial_5 - \frac{g_5}{\sqrt{2}} \Phi \right) \mathcal{H} \right\} + \rm{h.c.} \right].
\end{split}
\end{equation}
As mentioned in Section~\ref{sec:intro}, in the presence of SUSY, we need tree-level contributions from charged matter to the effective potential of $A_5$, which can be achieved by the charged matter taking non-zero VEVs at the boundaries. Such VEVs break gauge invariance, but this is allowed because we have already broken gauge invariance  by the Dirichlet boundary conditions for the boundary components of the gauge fields. 
These VEVs can be achieved by adding the following boundary-localized superpotential terms to the action:
\begin{equation}
\label{eq:singleaxion_SUGRA_S5_boundary_terms}
\delta \mathcal{S}_5 =  \int d^4x \int_0^L dx_5 \left[   \int d^2 \theta \left \{ \lambda^\prime (\mathcal{H} - v^\prime)^2 \ \delta (x_5) +  \lambda (\mathcal{H} - v)^2 \ \delta (x_5 - L) \right \} + \rm{h.c.} \right].
\end{equation}

Consider Dirichlet boundary conditions for $\mathcal{V}$ and $\mathcal{H}^c$ and Neumann boundary conditions for $\Phi$ and $\mathcal{H}$, at both the boundaries. We implement these boundary conditions via realizing the extra dimension with an interval as an ``orbifold'' of the circle. With the angular coordinate $(\theta)$ on the circle going from $-\pi$ to $\pi$, we identify the points $\theta$ with $-\theta$. Thus, half of the circumference of the extra-dimensional circle is the physical interval with $x_5$ going from $0$ to $\pi R \equiv L$, where $R$ is the radius of the circle. The Dirichlet and Neumann boundary conditions for the fields in an interval can be implemented by assigning, respectively, odd and even parity under orbifold $(\theta \rightarrow -\theta)$. (See \cite{Sundrum:2005jf} for a review of this.) 

Let us solve for the classical potential of this model. We need to integrate out the heavy fields (i.e. $\mathcal{H}, \mathcal{H}^c$ and the KK modes in $\mathcal{V}$) at tree-level to get an effective theory in terms of $\Phi$. We search for a supersymmetric vacuum of the full theory where inflation happens at an excited state with $\cancel{\textrm{SUSY}}$ vacuum energy $V_{\rm inf}$. Considering the inflationary energy scale to be much less than the masses of the heavy fields $\left(V_{\rm inf}^{1/4} \ll m_{KK}, m\right)$
\footnote{As can be seen in Section~\ref{subsec:inflation_trajectory}, $\frac{V_{\rm inf}^{1/4}}{m_{KK}} \sim \frac{v L}{\sqrt{f}} e^{-mL} \sim \sqrt{g} \cdot e^{-mL} \cdot \left(v L^{3/2}\right) $ which is small due to the smallness of $e^{-mL}$ and the hypermultiplet boundary VEVs $\sim v$.}, to the leading order in $\frac{V_{\rm inf}^{1/4}}{m_{KK}}$, for the purpose of the dynamics of the heavy fields, their ground state can be approximated to be supersymmetric even during inflation. 
Thus, we can integrate them out by using their SUSY equations of motion.  

Firstly, we can set $\mathcal{V}$ to zero since it contains only heavy fields and with zero VEVs. $A_\mu$ in $\mathcal{V}$ cannot have non-zero VEV due to Lorentz invariance. The $D$-scalar in $\mathcal{V}$ is an order parameter for SUSY and hence $ \langle D \rangle = 0$ for SUSY ground state. Of course the fermions in $\mathcal{V}$ have vanishing VEVs. This leaves us with only the following terms in the action:
\begin{equation}
\label{eq:5Daction_before_integrating_out_H_Hc}
\begin{split}
\mathcal{S}_5 &= \int d^4x \int_0^L dx_5 \left[ \int d^4 \theta \left \{ \frac{1}{2} (\Phi + \bar{\Phi})^2 +  \bar{\mathcal{H}}^c \mathcal{H}^c + \bar{\mathcal{H}} \mathcal{H} \right\}  \right. \\
&+ \left. \int d^2 \theta \left \{ \mathcal{H}^c \left(m + \partial_5 - \frac{g_5}{\sqrt{2}}\Phi\right) \mathcal{H} + \lambda^\prime (\mathcal{H} - v^\prime)^2 \ \delta (x_5) +  \lambda (\mathcal{H} - v)^2 \ \delta (x_5 - L) \right \} + \rm{h.c.} \right].
\end{split}
\end{equation}
The heavy charged matter fields $\mathcal{H}$ and $\mathcal{H}^c$, with 5D bulk masses $m \gtrsim m_{KK}$, can now be integrated out by imposing the following SUSY constraints:
\begin{equation}
\label{eq:SUSY_constraints_on_hypermultiplet}
\frac{\partial W}{\partial \mathcal{H}} = 0 = \frac{\partial W}{\partial \mathcal{H}^c}.
\end{equation}
Thus, we obtain the following 4D effective action\footnote{Here, all the 4D fields are in canonical normalization. The 4D gauge coupling $g$ is defined as: $\frac{1}{g^2} = \frac{L}{g_5^2}$.} for $\Phi$,
\begin{equation}
\label{eq:singleaxion_SUGRA_L4}
\begin{split}
\mathcal{S}_4 = &\int d^4x \left[ \int d^4 \theta \ \frac{1}{2} (\Phi + \bar{\Phi})^2 \right. \\
+ & \left. \int d^2 \theta \left( W_0 + \lambda \frac{v^2 \ e^{-mL} \ e^{\frac{gL}{\sqrt{2}}\Phi} + v^{\prime 2} \ e^{mL} \ e^{-\frac{gL}{\sqrt{2}}\Phi} - 2 v v^\prime}{e^{mL} \ e^{-\frac{gL}{\sqrt{2}}\Phi} + e^{-mL} \ e^{\frac{gL}{\sqrt{2}}\Phi}} + \rm{h.c.} \right) \right],
\end{split}
\end{equation}
where we take $\lambda = \lambda^\prime$ for technical simplicity. The derivatives on $\mathcal{H}$ and $\mathcal{H}^c$ at the boundaries in \eqref{eq:5Daction_before_integrating_out_H_Hc} are evaluated by taking into account their orbifold parity (even and odd, respectively). 

The K$\ddot{\mathrm{a}}$hler potential in \eqref{eq:singleaxion_SUGRA_L4} displays shift symmetry for $A_5$, which is the imaginary part of the scalar component of $\Phi$. However, integrating out the charged hypermultiplet using \eqref{eq:SUSY_constraints_on_hypermultiplet} also generates shift symmetry violating terms in the K$\ddot{\mathrm{a}}$hler potential. These corrections are however functions of $gL \Phi$ and suppressed by $e^{-mL}$, our modest expansion parameter. Thus, they contribute to the scalar potential only with $\frac{\Phi}{f}$-dependence $\left(f \equiv \frac{1}{gL}\right)$, not changing its qualitative form. Furthermore, the $e^{-mL}$ suppression makes these corrections sub-dominant and hence we neglect them here. 


The superpotential is the source of shift symmetry breaking for $A_5$ which is naturally suppressed by $e^{-mL}$ for $mL \gtrsim 1$ (see e.g. for $v \sim v^\prime$). This is a generic feature of extranatural inflation scenario where the compact extra dimension effectively acts as a ``filter'' for any far-UV physics by suppressing its contribution by $e^{- M_{\rm UV} L}$.
$W_0$ is a constant term in the superpotential which is relevant only in the presence of gravity, as we will see in Section~\ref{subsec:adding_sugra}.

\subsection{Bi-axion generalization to realize $f_{\rm eff} > M_{\rm Pl}$}
\label{subsec:biaxion_setup}

\begin{figure}[t]
\centering
\begin{subfigure}{.5\textwidth}
  \centering
  \includegraphics[width=0.9\textwidth]{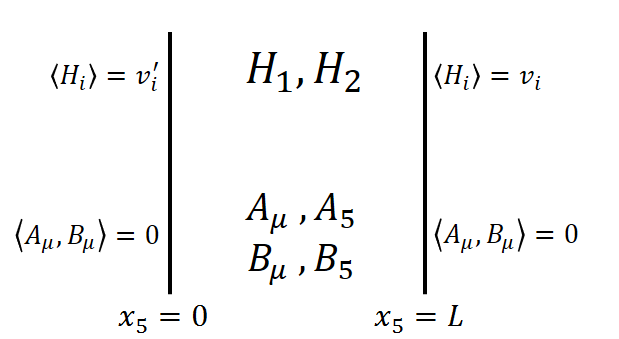}
  \caption{Non-SUSY}
  \label{fig:bi_axion_setup_nonSUSY}
\end{subfigure}%
\begin{subfigure}{.5\textwidth}
  \centering
  \includegraphics[width=0.9\textwidth]{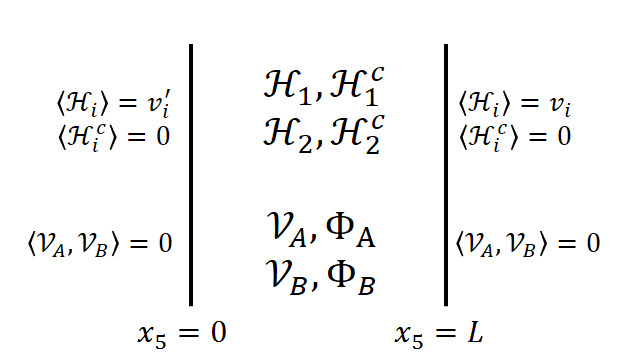}
  \caption{SUSY}
  \label{fig:bi_axion_setup_SUSY}
\end{subfigure}
\caption{Bi-axion inflation field content: (a) non-SUSY and (b) SUSY version. See text for details.}
\label{fig:bi_axion_setup_nonSUSY_SUSY}
\end{figure}

As mentioned in Section~\ref{sec:intro}, in order to have $f_{\rm eff} > M_{\rm Pl}$, we need to introduce two axions in such a way that one of their linear combinations has an effective super-Planckian field range. The non-SUSY version of bi-axion inflation has the setup as shown in Figure~\ref{fig:bi_axion_setup_nonSUSY}. There are two gauge fields $(A_M, B_M)$ with only $(A_5, B_5)$ having zero modes (by suitably assigning boundary conditions). The scalar fields $H_1$ and $H_2$ are charged under the gauge groups as $(0,1)$ and $(1, N)$, respectively. This field content can now be embedded into the respective 5D SUSY multiplets, as shown in Figure~\ref{fig:bi_axion_setup_SUSY}. By extending the construction from Section~\ref{subsec:5DSUSY_gaugetheory}, the full 5D action in this case can be obtained as follows:
\begin{equation}
\label{eq:biaxion_setup_S5}
\begin{split}
\mathcal{S}_5 =&  \int d^4x \int_0^L dx_5 \left[ 
\int d^2 \theta \left( \frac{1}{4} \mathcal{W}_{A, \alpha}^2 + \frac{1}{4} \mathcal{W}_{B, \alpha}^2 \right) + \rm{h.c.} \right.\\
+& \int d^4 \theta \left\{ \partial_5 \mathcal{V}_A - \frac{1}{\sqrt{2}} (\Phi_A + \bar{\Phi}_A) \right\}^2 
+ \left\{ \partial_5 \mathcal{V}_B - \frac{1}{\sqrt{2}} (\Phi_B + \bar{\Phi}_B) \right\}^2 \\
+& \int d^4 \theta \left\{ \left( \mathcal{H}_1^c e^{g_5 \mathcal{V}_B} \bar{\mathcal{H}_1^c} + \bar{\mathcal{H}_1} e^{-g_5 \mathcal{V}_B} \mathcal{H}_1 \right)
+ \left( \mathcal{H}_2^c e^{g_5 \left(\mathcal{V}_A + N \mathcal{V}_B \right)} \bar{\mathcal{H}_2^c} + \bar{\mathcal{H}_2} e^{-g_5 \left(\mathcal{V}_A + N \mathcal{V}_B \right)} \mathcal{H}_2 \right) \right\} \\
+& \int d^2 \theta \left\{ \mathcal{H}_1^c \left(m + \partial_5 - \frac{g_5}{\sqrt{2}} \Phi_B \right) \mathcal{H}_1
+ \mathcal{H}_2^c \left(m + \partial_5 - \frac{g_5}{\sqrt{2}} \left(\Phi_A + N \Phi_B \right) \right) \mathcal{H}_2 \right\} + \rm{h.c.} \\
+& \int d^2 \theta \left \{ \lambda_1^\prime (\mathcal{H}_1 - v_1^\prime)^2 \ \delta (x_5) +  \lambda_1 (\mathcal{H}_1 - v_1)^2 \ \delta (x_5 - L) \right \} + \rm{h.c.} \\
+& \left. \int d^2 \theta \left \{ \lambda_2^\prime (\mathcal{H}_2 - v_2^\prime)^2 \ \delta (x_5) +  \lambda_2 (\mathcal{H}_2 - v_2)^2 \ \delta (x_5 - L) \right \} + \rm{h.c.} \right].
\end{split}
\end{equation}

Similarly to how \eqref{eq:singleaxion_SUGRA_L4} was obtained starting from \eqref{eq:singleaxion_SUGRA_S5} and \eqref{eq:singleaxion_SUGRA_S5_boundary_terms} in the previous section, we can obtain the 4D effective K$\ddot{\mathrm{a}}$hler potential and superpotential for the two axion supermultiplets $(\Phi_A, \Phi_B)$ as follows:
\begin{equation}
\label{eq:biaxion_setup}
\begin{split}
K &= \frac{1}{2} (\Phi_A + \bar{\Phi}_A)^2 + \frac{1}{2} (\Phi_B + \bar{\Phi}_B)^2 , \\
W &= W_0 
+ \lambda_1 \frac{v_1^2 \ e^{-mL} \ e^{\frac{gL}{\sqrt{2}}(\Phi_A + N \Phi_B)} + v_1^{\prime 2} \ e^{mL} \ e^{-\frac{gL}{\sqrt{2}}(\Phi_A + N \Phi_B)} - 2 v_1 v_1^\prime}{e^{mL} \ e^{-\frac{gL}{\sqrt{2}}(\Phi_A + N \Phi_B)} + e^{-mL} \ e^{\frac{gL}{\sqrt{2}}(\Phi_A + N \Phi_B)}} \\
 &+ \lambda_2 \frac{v_2^2 \ e^{-mL} \ e^{\frac{gL}{\sqrt{2}}\Phi_B} + v_2^{\prime 2} \ e^{mL} \ e^{-\frac{gL}{\sqrt{2}}\Phi_B} - 2 v_2 v_2^\prime}{e^{mL} \ e^{-\frac{gL}{\sqrt{2}}\Phi_B} + e^{-mL} \ e^{\frac{gL}{\sqrt{2}}\Phi_B}}.
\end{split}
\end{equation} 
We would like to highlight here that in \eqref{eq:biaxion_setup_S5}, and hence also in \eqref{eq:biaxion_setup}, all the scales and field ranges are sub-Planckian.

\subsection{Adding SUGRA and identifying the SUSY vacuum}
\label{subsec:adding_sugra}

We have not considered the effects of gravity so far in obtaining the $\mathcal{L}_{4, \rm eff} (\Phi_A, \Phi_B)$ of \eqref{eq:biaxion_setup}. But now we can use this 4D effective $K$ and $W$ to compute the SUGRA scalar potential $(V_{\rm SUGRA})$ directly in 4D (see Appendix~\ref{sec:appendix_sugra_basics}). With this strategy, from effective field theory perspective, we could only be missing $M_{\rm Pl}$-suppressed terms e.g. $K \ni (\Phi_A + \bar{\Phi}_A)^4 , (\Phi_B + \bar{\Phi}_B)^4$. In the case of SUSY bi-axion model, as highlighted below \eqref{eq:biaxion_setup}, the range (and hence also the VEVs) of fields in $\Phi_A$ and $\Phi_B$ is sub-Planckian, thus making the above-mentioned $M_{\rm Pl}$-suppressed terms also sub-dominant. We would like to highlight here that in the case of a single axion (in Section~\ref{subsec:5DSUSY_gaugetheory}), such $M_{\rm Pl}$-suppressed terms in \eqref{eq:singleaxion_SUGRA_L4} are \textit{not} sub-dominant due to the super-Planckian range of the fields. Hence the truncation of the $\Phi/M_{\rm Pl}$ expansion is uncontrolled in this case.
In Section~\ref{sec:observable_signals}, we will see that higher order K$\ddot{\mathrm{a}}$hler interactions can have interesting observable effects if they are stronger than $M_{\rm Pl}$-suppressed.

The $W_0$ parameter in \eqref{eq:biaxion_setup} is now physical, due to the presence of gravity, and it will contribute to the vacuum energy. We will consider a boundary-localized contribution to $W_0$ such that the net post-inflationary vacuum energy is (approximately) zero.

In order for the inflationary picture to be compatible with low energy SUSY (broken only at a scale somewhat above the EW scale) and approximately zero cosmological constant as observed today, the vacuum of post-inflationary dynamics should be SUSY-preserving and with zero vacuum energy. Thus, it seems that the inflation endpoint has to (approximately) satisfy the following three conditions: (1) unbroken SUSY $(\left\langle D_{\Phi_i} W \right\rangle =  0)$, (2) zero vacuum energy $(\langle W \rangle = 0)$, and (3) local minimum\footnote{Global minimum can be separated enough in the field space from this local minimum such that the decay via tunneling does not happen even on the cosmological timescales.} of $V_{\rm SUGRA}$. However, as shown below, (1) and (2) automatically imply (3), i.e. a point in the field space satisfying $D_{\Phi_i} W = 0$ and $W = 0$ implies that it is automatically at a local minimum of $V_{\rm SUGRA}$, so we do not bother to check (3) further.

Consider, for simplicity, a single chiral superfield $\Phi$ for which $V_{\rm SUGRA}$ is
\begin{equation}
V = e^K \left( K^{-1}_{\Phi \bar{\Phi}} \ \left| D_{\Phi} W \right|^2  - 3 \ |W|^2 \right).
\end{equation}
Now, for $D_{\Phi} W = 0$ and $W = 0$, one can clearly see that,
\begin{equation}
\partial_{\Phi} V = 0 = \partial_{\bar{\Phi}} V, \ 
\partial_{\bar{\Phi}} \partial_{\Phi} V = e^K K^{-1}_{\Phi \bar{\Phi}} \ \left| \partial_\Phi D_\Phi W \right|^2 , \
\partial_\Phi^2 V = 0 = \partial_{\bar{\Phi}}^2 V,
\end{equation}
and for $\Phi = (\eta + i \phi)/\sqrt{2}$,
\begin{equation}
\label{eq:local_minimum_for_VSUGRA}
\partial_{\eta} V = 0 = \partial_{\phi} V, \ \partial_\eta^2 V = \partial_\phi^2 V = \frac{1}{2} e^K K^{-1}_{\Phi \bar{\Phi}} \ \left| \partial_\Phi D_\Phi W \right|^2.
\end{equation}
Thus, for $K^{-1}_{\Phi \bar{\Phi}} \geq 0$ and $K_{\bar{\Phi} \Phi \Phi}, K_{\bar{\Phi} \Phi \bar{\Phi} \Phi}$ finite, \eqref{eq:local_minimum_for_VSUGRA} implies a local minimum of $V_{\rm SUGRA}$. The same proof can be applied for multiple chiral superfields $\Phi_i$, with $K^{-1}_{\Phi_i \bar{\Phi}_i} \geq 0$ in the mass basis and no singularities in higher derivatives of $K$. These conditions are satisfied in our cases of interest, since we mostly have $K^{-1}_{\Phi_i \bar{\Phi}_i} \approx 1$ (see \eqref{eq:biaxion_setup}) with corrections suppressed by high scales $\Lambda$ and small field VEVs (see Section~\ref{subsec:primordial_NG}).

Furthermore, the conditions $D_{\Phi_i} W = 0$ and $W = 0$ are equivalent to the conditions 
\begin{equation}
\partial_{\Phi_i} W = 0 \ , \ W = 0,
\end{equation}
since $D_{\Phi_i} W = \partial_{\Phi_i} W + \left(\partial_{\Phi_i} K \right) W$. This hugely simplifies identifying the inflation endpoint analytically. The conditions $\partial_{\Phi_A} W = 0 = \partial_{\Phi_B} W$ can be satisfied for the superpotential in \eqref{eq:biaxion_setup} minimally by the following choice for the parameters\footnote{The simplest choice with $v_1 = v_1^\prime = v_2 = v_2^\prime$ does not admit a solution to $\partial_{\Phi_A} W = 0 = \partial_{\Phi_B} W$ when restricted to sub-Planckian field values.} that govern the hypermultiplet VEVs at the boundaries (see \eqref{eq:biaxion_setup_S5})
: 
\begin{equation}
\label{eq:boundary_VEVs_for_hypermultiplets}
v_1 = v_1^\prime = v_2 \equiv v , v_2^\prime \sim v e^{-mL}.
\end{equation}
In order to avoid having significant fine-tuning for choosing $v_2^\prime \sim v e^{-mL}$, we consider $e^{-mL} \sim \mathcal{O}(1)$, while still having $e^{-mL} < 1$ for valid perturbative expansion (e.g. $e^{-mL} \approx 1/3$ with $mL \approx 1.1$). With the choice of parameters $v_i, v_i^\prime$ as in \eqref{eq:boundary_VEVs_for_hypermultiplets}, and after doing a change of basis from $\left( \Phi_A, \Phi_B \right)$ to $\left( \Phi_h, \Phi_l \right)$ as defined by
\begin{equation}
\label{eq:BasisChange_Phi_Phitilde_to_Phil_Phih}
\Phi_h \equiv \Phi_B + \frac{1}{N} \Phi_A \ , \ \Phi_l \equiv \Phi_A - \frac{1}{N} \Phi_B,
\end{equation}
the superpotential from \eqref{eq:biaxion_setup} becomes
\begin{equation}
\label{eq:biaxion_SUGRA_superpotential_withBoundaryVEVchoice}
\begin{split}
\frac{W}{v^2} &= \frac{W_0}{v^2} + \lambda_1 \left[1 - \frac{1}{\cosh{\left( mL -\frac{gL}{\sqrt{2}} N \Phi_h \right)}} \right] \\
&- 2 \lambda_2 \ e^{-2mL} \ e^{\frac{gL}{\sqrt{2}}\left(\Phi_h - \frac{\Phi_l}{N} \right)} \left[1 - \cosh{\left(\frac{gL}{\sqrt{2}} \left(\Phi_h - \frac{\Phi_l}{N} \right) \right)}  \right] + \mathcal{O}\left(e^{-4mL}  \right).
\end{split}
\end{equation}
%
%

Now, we can identify the required Minkowski SUSY endpoint of inflation. Firstly, we identify VEVs of all the scalars, $\Phi_k = \frac{1}{\sqrt{2}} (\eta_k + i \phi_k)$, at inflation end by solving $\partial_{\Phi_k} W = 0$ to obtain
\begin{equation}
\label{eq:inflation_endpoint}
\langle \phi_l \rangle = 0 = \langle \phi_h \rangle \ , \ \langle \eta_h \rangle \approx \frac{f m L}{N} \ , \ \langle \eta_l \rangle \approx f m L,
\end{equation}
with $f \equiv \frac{2}{g L}$. Then, plugging these VEVs back into \eqref{eq:biaxion_SUGRA_superpotential_withBoundaryVEVchoice}, we can enforce $\langle W  \rangle = 0$. This self-consistently demands $W_0$ to be chosen to cancel the terms in \eqref{eq:biaxion_SUGRA_superpotential_withBoundaryVEVchoice} sub-dominant in $e^{-mL}$, i.e.
\begin{equation}
\label{eq:W0_choice}
W_0 \sim v^2 \cdot \mathcal{O} \left(e^{-4 m L}\right),
\end{equation}
where, as mentioned below \eqref{eq:boundary_VEVs_for_hypermultiplets}, $e^{-mL}$ is our modest expansion parameter.

One can clearly see from \eqref{eq:biaxion_SUGRA_superpotential_withBoundaryVEVchoice} that $W \approx W\left(N \Phi_h, \Phi_l/N \right)$, for $N \gg 1$, and hence the scalar potential will be of the form
\begin{equation}
\label{eq:VSUGRA_scaling}
V \approx V\left( \frac{N \eta_h}{f},\frac{N \phi_h}{f}, \eta_l, \frac{\phi_l}{N f} \right).
\end{equation}
Due to the $e^K$ contribution from $V_{\rm SUGRA}$, and that $K \ni \frac{1}{2} \left(\Phi_l + \bar{\Phi}_l \right)^2$ has $\eta_l$- but no $\phi_l$-dependence, the potential along $\eta_l$ varies over $M_{\rm Pl}$, and not $N f > M_{\rm Pl}$. 
As we will detail in Sec.~\ref{subsec:inflation_trajectory}, from \eqref{eq:VSUGRA_scaling} we can power-count $m_{\eta_h}, m_{\phi_h} \sim H_{\rm inf} \cdot \mathcal{O}\left(\frac{N^2}{f}\right) \gg H_{\rm inf}$ and $m_{\eta_l} \sim \mathcal{O}(H_{\rm inf})$, while only $m_{\phi_l} \ll H_{\rm inf}$.
We will show in the Sec.~\ref{subsec:inflation_trajectory} that after integrating out the heavy fields ($\eta_h, \phi_h, \eta_l$), we get $V_{\rm eff}\left(\frac{\phi_l}{N f} \right)$ such that $\phi_l$ has an effective super-Planckian field range: $f_{\rm eff} = N f > M_{\rm Pl}$ with $f < M_{\rm Pl}$ and $N \gg 1$. Thus, we expect that the SUSY vacuum $\phi_l = 0$ can be approached from some $\phi_l^{\rm initial} \sim \mathcal{O}(Nf)$ along a slow-roll potential with the slow-roll parameters $\epsilon, \eta \sim \left(\frac{M_{\rm Pl}}{Nf}\right)^2 \ll 1$.

As in the case of single axion supermultiplet (see below \eqref{eq:singleaxion_SUGRA_L4}), integrating out the charged hypermultiplets in \eqref{eq:biaxion_setup_S5} also generates shift symmetry violating terms in the K$\ddot{\mathrm{a}}$hler potential in \eqref{eq:biaxion_setup}. Similar to the case of single axion supermultiplet, these corrections are functions of $\frac{\Phi_l}{f_{\rm eff}}$ which maintain the form of the effective inflationary potential, $V_{\rm eff}\left(\frac{\phi_l}{f_{\rm eff}}\right)$, i.e. effective super-Planckian field range for $\phi_l$. Furthermore, they are suppressed by powers of $e^{-mL}$, our modest expansion parameter, which makes them sub-dominant and hence we neglect them here.


\section{Inflationary history}
\label{sec:inflationary_history}

In this section, we describe various aspects of the inflationary history from our SUSY bi-axion model. Here we discuss the inflationary trajectory, SUSY breaking during inflation, accounting for the SUSY breaking vacuum that we occupy today, and the interplay of different fine-tunings within this model. 

\subsection{Inflationary trajectory}
\label{subsec:inflation_trajectory}

In order to identify the inflationary trajectory and the effective potential along it, we first consider the general problem where a potential depends on some heavy fields $\vec{H}$ and some light fields $\vec{L}$. We expand the potential to quadratic order in $\vec{H}$ while keeping it to all orders in $\vec{L}$: 
\begin{equation}
\label{eq:general_potential_quadratic_in_heavies}
V\left(\vec{H}, \vec{L}\right) = V\left(\langle\vec{H}\rangle, \vec{L}\right) + A_i\left(\vec{L} \right) \cdot \delta H_i + \frac{1}{2}m^2_{ij}\left(\vec{L} \right) \cdot \delta H_i \cdot \delta H_j + \mathcal{O}(\delta H^3).
\end{equation}
Here, $\langle \vec{H} \rangle$ are VEVs of the heavy fields at the post-inflationary SUSY vacuum, while $\delta \vec{H}$ are the fluctuations away from $\langle \vec{H} \rangle$ in the course of inflation.
The expansion coefficients are
\begin{equation}
\label{eq:tadpole_massmatrix_heavies_during_inflation}
A_i\left(\vec{L} \right) \equiv \partial_{H_i} V\left(\langle\vec{H}\rangle, \vec{L}\right) \ , 
\ m^2_{ij}\left(\vec{L} \right) \equiv \partial_{H_i} \partial_{H_j} V\left(\langle\vec{H}\rangle, \vec{L}\right),
\end{equation}
which are functions of $\vec{L}$. We can now integrate out the heavy fields by extremizing \eqref{eq:general_potential_quadratic_in_heavies} with respect to the heavy fluctuations $\delta\vec{H}$, for given light fields $\vec{L}$,
thereby getting an effective potential for $\vec{L}$ as
\begin{equation}
\label{eq:general_effective_potential_for_light_fields_after_integrating_out_the_heavies}
V_{\rm eff}\left(\vec{L} \right) \approx V\left(\langle\vec{H}\rangle, \vec{L}\right) + \frac{1}{2} \left[ A_i \left(m^2\right)^{-1}_{ij} A_j    \right] \left(\vec{L}\right).
\end{equation}

In the case of our SUSY bi-axion inflation model, the heavy and light fields along the inflationary trajectory are, respectively, 
\begin{equation}
\label{eq:heavy_light_fields_during_inflation}
\vec{H} = (\eta_h, \phi_h, \eta_l) \ , \ \vec{L} = (\phi_l).
\end{equation}
The coefficients $A_i$ and mass matrix $m^2_{ij}$ in $V_{\rm SUGRA}$ \eqref{eq:appendix_V_SUGRA}, for the superpotential \eqref{eq:biaxion_SUGRA_superpotential_withBoundaryVEVchoice}, scale parametrically as follows:
\begin{equation}
\label{eq:going_rate_of_A_massmatrix_of_heavies_during_inflation}
\begin{split}
A_i\left(\vec{L}\right) &\sim \left( N^2 , N^2 , \frac{1}{N} \right) \frac{V\left(\langle\vec{H}\rangle, \vec{L}\right)}{f} , \\
m^2_{ij}\left(\vec{L}\right) &\sim \begin{pmatrix} N^4 & N^2 & N \\ N^2 & N^4 & 0 \\ N & 0 & f^2 \end{pmatrix}  \frac{V\left(\langle\vec{H}\rangle, \vec{L}\right)}{f^2},
\end{split}
\end{equation}
where the indices $i, j$ run over $\vec{H}$ in the same order as in \eqref{eq:heavy_light_fields_during_inflation}. We can now estimate the parametric size of the heavy fluctuations during inflation as
\begin{equation}
\left\langle \delta \vec{H} \right\rangle \sim \left(\frac{f}{N^2} , \frac{f}{N^2} , \frac{1}{Nf} \right).
\end{equation}
This then implies that the $\mathcal{O}(\delta H^3)$ term that we dropped in \eqref{eq:general_potential_quadratic_in_heavies} is sub-dominant, suppressed by the small parameters $\frac{1}{N}$ and $\frac{1}{Nf}$, and hence can be ignored along the inflationary trajectory. 

\begin{figure}[t]
\centering
\includegraphics[width=0.9\textwidth]{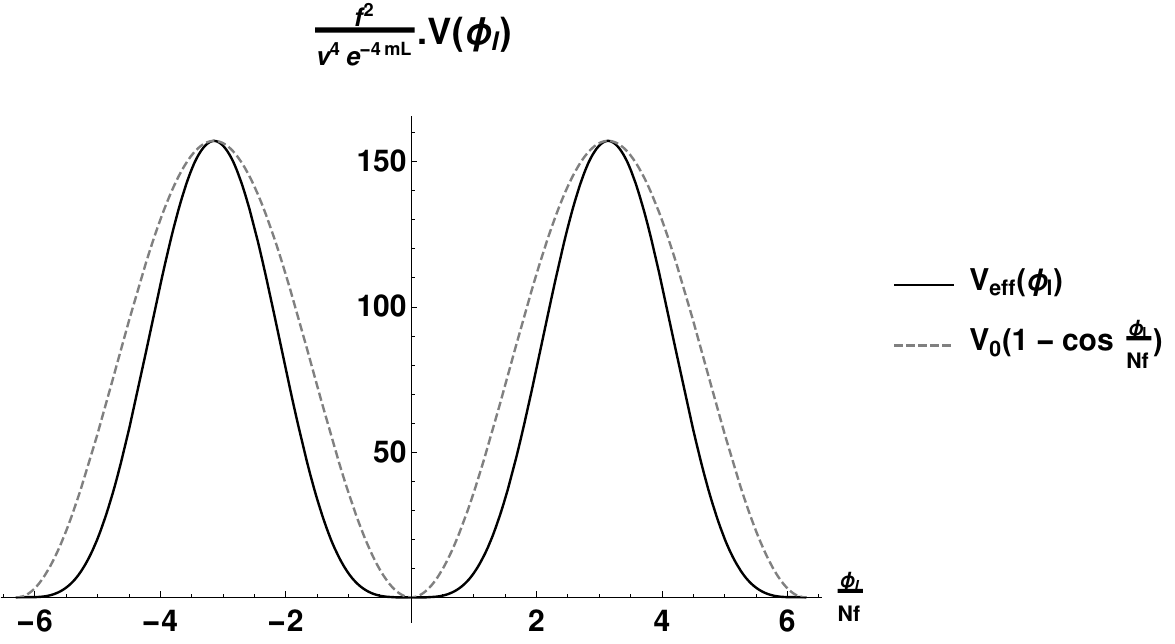}
\caption{The dark line refers to the effective inflationary potential $V_{\rm eff}(\phi_l)$ after numerically integrating out the heavy fields $\eta_h, \phi_h, \eta_l$ all along the inflationary trajectory. For comparison, a pure cosine potential with the magnitude matching to that of $V_{\rm eff}(\phi_l)$ is plotted as the dashed line. Inflation can start close to the hilltop of $V_{\rm eff}(\phi_l)$, with $V_{\rm inf} \sim \frac{v^4 e^{-4mL}}{f^2}$.}
\label{fig:Veff_phil_numerical}
\end{figure}

The mass eigenvalues of the heavy fluctuations $\delta\vec{H}$ are as follows:
\begin{equation}
\label{eq:mass_eigenvalues_of_heavies_during_inflation}
\begin{split}
m_{1,2}^2 &\approx m_{\eta_h, \phi_h}^2 \sim N^4 \frac{V\left(\langle\vec{H}\rangle, \langle\vec{L}\rangle\right)}{f^2} \sim \frac{N^4}{f^2} H_{\rm inf}^2 , \\ 
m_3^2 &\approx m_{\eta_l}^2 \sim V\left(\langle\vec{H}\rangle, \vec{L}\right) \sim H_{\rm inf}^2\left(\vec{L}\right).
\end{split}
\end{equation}
Thus, the heavy mass-squared eigenvalues are all positive and much larger than $m_{\phi_l}^2 \sim \frac{V\left(\langle\vec{H}\rangle, \vec{L}\right)}{N^2 f^2}$. Hence, we can integrate out the heavy fluctuations $\delta\vec{H}$ all along the inflationary trajectory yielding
\begin{equation}
\label{eq:Veff_phil_after_integrating_out_heavies}
V_{\rm eff}\left(\phi_l\right) \approx V\left(\langle\vec{H}\rangle, \frac{\phi_l}{Nf}\right) \cdot \left[1 + \mathcal{O}(1) + \mathcal{O}\left(\frac{1}{N^2 f^2}\right) \right].
\end{equation} 
Here, the second term comes from integrating out the mass eigenstates $\left(H_1 , H_2 \right) \approx \left(\eta_h, \phi_h \right)$ while the third term comes from integrating out $H_3 \approx \eta_l$. Since the contribution from $\left(H_1 , H_2 \right)$ is of the same order as $V\left(\langle\vec{H}\rangle, \frac{\phi_l}{Nf}\right)$, we perform the integration out of $\delta \vec{H}$ numerically. Firstly, we verify that as suggested by the parametric estimates in \eqref{eq:going_rate_of_A_massmatrix_of_heavies_during_inflation} and \eqref{eq:mass_eigenvalues_of_heavies_during_inflation}, the heavy mass-squared eigenvalues are indeed much larger than $m_{\phi_l}^2$ all along the inflationary trajectory. The numerically computed $V_{\rm eff}(\phi_l)$ is as shown in Fig.~\ref{fig:Veff_phil_numerical}.
It has the following approximate analytic form,
\begin{equation}
V_{\rm eff}(\phi_l) \sim H_{\rm inf}^2 \left( 1 - \cos \frac{\phi_l}{N f} \right) \ , \ 
        H_{\rm inf} \sim \lambda_2 \frac{v^2}{f} \ e^{-2 m L},
\end{equation}
which is the leading contribution in terms of the small parameters $1/N$, $f$ and $e^{-mL}$. Thus, the SUSY bi-axion model effectively provides an approximate Natural Inflation model with $f_{\rm eff} = N f > M_{\rm Pl}$, where the best-fit values are \cite{Akrami:2018odb}
\begin{equation}
\label{eq:natural_inflation_bestfit_values}
f_{\rm eff} = N f \sim 10 M_{\rm Pl} \ , \  V_{\rm inf}^{1/4} \sim 10^{16} GeV.
\end{equation}
The precise inflationary potential, $V_{\rm eff}(\phi_l)$, does contain ``higher harmonics'' in $\frac{\phi_l}{Nf}$. Although, these do not affect the qualitative features as outlined above, they can play an important role in precision fitting to the CMB data which we will explore in a future work.

As can be seen from \eqref{eq:mass_eigenvalues_of_heavies_during_inflation}, the heavy fields $\eta_h , \phi_h$ are much heavier than $H_{\rm inf}$:
\begin{equation}
\label{eq:heavy_fields_masses}
m_{\eta_h} , m_{\phi_h} \sim H_{\rm inf} \frac{N^2}{f}. 
\end{equation}
While sinflaton $(\eta_l)$, the real scalar partner of the inflaton, has an intermediate mass, 
\begin{equation}
m_{\eta_l} \sim \mathcal{O}(H_{\rm inf}), 
\end{equation}
about which we will discuss more in Section~\ref{subsubsec:NG_sinflaton}.

\subsection{SUSY breaking during inflation}
\label{subsec:SUSYbreaking_during_inflation}

As mentioned earlier in Section~\ref{sec:KL_toy_model}, the approximate de Sitter geometry during inflation requires it to be an excited state with spontaneous $\cancel{\textrm{SUSY}}$ on top of the post-inflationary SUSY vacuum. The SUSY order parameters (see Appendix~\ref{sec:appendix_sugra_basics}) for $\Phi_h$ and $\Phi_l$ evaluated along the inflationary trajectory are
\begin{equation}
\label{eq:SUSY_order_param_during inflation}
    \begin{split}
        D_{\Phi_h} W \big \rvert_{\langle \Phi_h \rangle , \langle \eta_l \rangle} 
        &\sim \lambda_2 \frac{v^2}{f} e^{-2m L} \ e^{-i \frac{\phi_l}{Nf}} \left(1 - \cos{\frac{\phi_l}{Nf}} \right), \\
        D_{\Phi_l} W \big \rvert_{\langle \Phi_h \rangle , \langle \eta_l \rangle} 
        &\sim \mathcal{O}\left(\frac{1}{N}\right) \cdot D_{\Phi_h} W \big \rvert_{\langle \Phi_h \rangle , \langle \eta_l \rangle}.
    \end{split}
\end{equation}
Here, we can clearly see that the SUSY order parameters are zero only at the vacuum (i.e. $\phi_l = 0$).

As described in Appendix~\ref{sec:appendix_sugra_basics}, the massless Goldstino of spontaneous $\cancel{\textrm{SUSY}}$, which is ``eaten'' by gravitino to become massive, is given by the linear combination of all the fermions weighted by the respective SUSY order parameters: $\psi_{\textrm{Goldstino}} \propto \langle D_{\Phi_i} W \rangle \cdot \psi_i$. Hence, the SUSY order parameters during inflation \eqref{eq:SUSY_order_param_during inflation} imply that,
\begin{equation}
\psi_{\textrm{Goldstino}} \sim \psi_h + \mathcal{O}\left(\frac{1}{N}\right) \cdot \psi_l,
\end{equation}
i.e. the Goldstino during inflation belongs mostly to the heavy sector $(\Phi_h)$ and not to the inflaton sector $(\Phi_l)$. In other words, $\cancel{\textrm{SUSY}}$ during inflation is caused mostly by the heavy sector $(\Phi_h)$. This feature of our SUSY bi-axion model is very similar to the KLR model discussed in Section~\ref{sec:KL_toy_model} where the Goldstino during inflation  belongs solely to the heavy sector $(S)$ (see \eqref{eq:SUSY_order_param_KL_model}).

\subsection{SUSY breaking after inflation}
\label{subsec:SUSYbreaking_after_inflation}

We need to account for the post-inflationary $\cancel{\textrm{SUSY}}$ vacuum that we occupy today which we have neglected so far. In this section, we look for any possibly significant effects of this $\cancel{\textrm{SUSY}}$ vacuum on the inflationary dynamics. 
We consider a boundary-localized Polonyi-type sector with 
\begin{equation}
\delta K_4 = \bar{X} X - \frac{(\bar{X} X)^2}{\Lambda^2} \ , \  \delta W_4 = \Lambda_{\cancel{\rm SUSY}}^2 X,
\end{equation}
which undergoes spontaneous $\cancel{\textrm{SUSY}}$ at a scale $\sim \Lambda_{\cancel{\rm SUSY}}$. Consider the $\cancel{\textrm{SUSY}}$ from this hidden sector to be minimally communicated to the Standard Model via $M_{\rm Pl}$ suppressed interactions which then implies an intermediate scale $\cancel{\textrm{SUSY}}$ $\Lambda_{\cancel{\rm SUSY}} \sim \sqrt{v_{\rm weak} \cdot M_{\rm Pl}} \sim 10^{11} GeV$. We expect that as long as $\Lambda_{\cancel{\rm SUSY}} \ll H_{\rm inf}$ or even $V_{\rm inf}^{1/4}$, the effect of this $\cancel{\textrm{SUSY}}$ $X$-sector on the inflationary dynamics will be negligible. (See also e.g. \citep{Dudas:2015lga}.) We show below that indeed this expectation is borne out here. 

The scalar field $X$ in the Polonyi sector gets $\mathcal{O}(H_{\rm inf})$ mass during inflation, similar to that of $\eta_l$. Thus, $X$ can be added to the ``heavy'' fields $\vec{H}$ of \eqref{eq:heavy_light_fields_during_inflation} and essentially the same procedure as described in Sec.~\ref{subsec:inflation_trajectory} can be repeated to integrate them out along the inflationary trajectory. The tadpole and mass matrix terms (see \eqref{eq:tadpole_massmatrix_heavies_during_inflation}) involving $X$ field scale as follows:
\begin{equation}
\label{eq:tadpole_massmatrix_X}
\begin{split}
A_X(\phi_l) &\sim \Lambda_{\cancel{\rm SUSY}}^2 H_{\rm inf}(\phi_l) \cdot f \ , \ m^2_{XX}(\phi_l) \sim H_{\rm inf}^2(\phi_l) , \\
m^2_{\eta_h X}(\phi_l) &\sim m^2_{\phi_h X}(\phi_l) \sim \Lambda_{\cancel{\rm SUSY}}^2 H_{\rm inf}(\phi_l) \cdot N \ , \  
m^2_{\eta_l X}(\phi_l) \sim \Lambda_{\cancel{\rm SUSY}}^2 H_{\rm inf}(\phi_l) \frac{1}{N}. 
\end{split}
\end{equation}
The fluctuations of $X$ during inflation can be estimated from \eqref{eq:tadpole_massmatrix_X} as 
\begin{equation}
\delta X \sim \frac{\Lambda_{\cancel{\rm SUSY}}^2 f}{V_{\rm inf}^{1/2}(\phi_l)}.
\end{equation}
The $\mathcal{O}(\delta H^3)$ terms in \eqref{eq:general_potential_quadratic_in_heavies} involving $X$ are hence sub-dominant, suppressed by $\frac{\Lambda_{\cancel{\rm SUSY}}}{V_{\rm inf}^{1/4}}$. We can now integrate out $X$ following \eqref{eq:general_effective_potential_for_light_fields_after_integrating_out_the_heavies} yielding
\begin{equation}
\delta V_{\rm eff}(\phi_l) \sim \Lambda_{\cancel{\rm SUSY}}^4 \cdot f^2 \cdot \left[ 1 + \frac{1}{N^2} \frac{\Lambda_{\cancel{\rm SUSY}}^4}{V_{\rm eff}(\phi_l)} \right] 
\end{equation}
This $\delta V_{\rm eff}(\phi_l)$ is much smaller than the $V_{\rm eff}(\phi_l)$ of \eqref{eq:Veff_phil_after_integrating_out_heavies}. Thus, as expected, for $\Lambda_{\cancel{\rm SUSY}} \sim 10^{11}$ GeV and $V_{\rm inf}^{1/4} \sim 10^{16}$ GeV (see \eqref{eq:natural_inflation_bestfit_values}) satisfying $\Lambda_{\cancel{\rm SUSY}} \ll V_{\rm inf}^{1/4}$, the $\cancel{\textrm{SUSY}}$ $X$-sector gives a negligible contribution to $V_{\rm eff}(\phi_l)$ and hence does not significantly affect the inflationary dynamics.

\subsection{Interplay of electroweak, cosmological constant and superpotential tunings}
\label{subsec:finetunings}

In order to have (almost) vanishing vacuum energy after the end of inflation, as discussed in Section~\ref{subsec:adding_sugra}, we need to have $\langle W \rangle = 0$ which can be achieved by tuning the $W_0$ parameter in \eqref{eq:biaxion_setup}. We also need to account for the SUSY breaking vacuum that we occupy today. Here, we evaluate this combined fine-tuning which displays an interesting interplay with the electroweak (EW) and cosmological constant (CC) fine-tunings.
 
Consider the $\cancel{\textrm{SUSY}}$ hidden sector of Sec.~\ref{subsec:SUSYbreaking_after_inflation} which minimally communicates to the Standard Model via gravity mediation (i.e. $M_{\rm Pl}$ suppressed interactions). In order for it to address the electroweak hierarchy problem, this requires that $\frac{V_{\cancel{\rm SUSY}}^{\rm today}}{M_{\rm Pl}^2} \sim v_{\rm weak}^2$.\footnote{ $V_{\cancel{\rm SUSY}}^{\rm today} = \Lambda_{\cancel{\rm SUSY}}^4$} This implies the following fine-tuning in the EW sector,
\begin{equation}
\label{eq:finetuning_EW}
T_{\rm EW} \sim \frac{v_{\rm weak}^2 M_{\rm Pl}^2}{V_{\cancel{\rm SUSY}}^{\rm today}},
\end{equation}
which can be minimized with $\left(V_{\cancel{\rm SUSY}}^{\rm today}\right)^{1/4} \sim \sqrt{v_{\rm weak} M_{\rm Pl}}$, as is standard. 
This $\cancel{\textrm{SUSY}}$ sector and also $\Delta W_0 \neq 0$ in \eqref{eq:biaxion_setup} give contributions to the CC today as below:
\begin{equation}
\label{eq:CC_contributions}
\textrm{CC} = - 3 \frac{\Delta W_0^2}{M_{\rm Pl}^2} + V_{\cancel{\rm SUSY}}^{\rm today} \stackrel{(\rm obs.)}{\sim} \textrm{meV}^4 .
\end{equation}
The two terms in the above equation have typical sizes of $\sim \frac{v^4}{M_{\rm Pl}^2}$ (see \eqref{eq:W0_choice}) and $\sim v_{\rm weak}^2 M_{\rm Pl}^2$, respectively, which consist of a priori different and unrelated scales. This implies that multiple contributions to $\Delta W_0$, each of magnitude $\sim \frac{v^2}{M_{\rm Pl}}$, must first cancel to within $\sqrt{V_{\cancel{\rm SUSY}}^{\rm today}}$. Hence, we have the following fine-tuning in the contributions to $W_0$:
\begin{equation}
\label{eq:finetuning_W0}
T_{W_0} \sim \frac{\sqrt{V_{\cancel{\rm SUSY}}^{\rm today}} M_{\rm Pl}}{v^2}.
\end{equation}   
Once the two terms on the right hand side of \eqref{eq:CC_contributions} are of the same order, they still have to cancel to give $\textrm{CC} \sim \textrm{meV}^4$ as observed today. This amounts to having the following usual CC fine-tuning:
\begin{equation}
\label{eq:finetuning_CC}
T_{\rm CC} \sim \frac{\textrm{meV}^4}{V_{\cancel{\rm SUSY}}^{\rm today}}.
\end{equation}

As can be seen from \eqref{eq:finetuning_EW} and \eqref{eq:finetuning_CC}, the EW and CC fine-tunings favor $\cancel{\textrm{SUSY}}$ at low-scale. However, \eqref{eq:finetuning_W0} shows that the $W_0$ fine-tuning displays preference for $\cancel{\textrm{SUSY}}$ at high-scale! But, the net fine-tuning, assuming that these three are independent of each other, is
\begin{equation}
\label{eq:finetuning_net}
T_{\rm net} = T_{\rm EW} \times T_{W_0} \times T_{CC} \sim \frac{v_{\rm weak}^2 M_{\rm Pl}^3 \textrm{meV}^4}{v^2} \left(V_{\cancel{\rm SUSY}}^{\rm today}\right)^{-3/2}. 
\end{equation}
This shows a net preference for $\cancel{\textrm{SUSY}}$ at low-scale, namely close to the EW scale. 

Our considerations here are reminiscent of comparable tuning issues that arise in high-scale string-derived SUGRA theories, in particular the necessary existence of a high-scale $W_0$ which makes the tuning worse. See \cite{Douglas:2006es} for a review. For a sample choice of the parameters, $V_{\cancel{\rm SUSY}}^{\rm today} \sim v_{\rm weak}^2 M_{\rm Pl}^2$ and $v^2 \sim (0.1 M_{\rm Pl})^3$, we see that the net tuning in \eqref{eq:finetuning_net} is considerable $\left(T_{\rm net} \sim 10^{-100}\right)$, predominantly because of the Cosmological Constant Problem. However, such a residual tuning is still acceptable in the context of the anthropic principle or some as yet unknown mechanism solving this problem. See \cite{Weinberg:1988cp, Bousso:2007gp} for a review. 

\section{Observable signals}
\label{sec:observable_signals}

In this section we discuss the phenomenology of our SUSY bi-axion model. The observable signals from this model can come in the form of primordial non-Gaussianities mediated by heavy particles, sinflaton being the prime candidate for this. Also, ``higher harmonic'' terms in the inflaton potential can give rise to periodic modulations in the CMB.

\subsection{Primordial non-Gaussianities}
\label{subsec:primordial_NG}

As first introduced in \cite{Chen:2009we} and further illustrated in \cite{Chen:2009zp, Chen:2010xka, Baumann:2011nk, Assassi:2012zq, Chen:2012ge, Pi:2012gf, Noumi:2012vr, Arkani-Hamed:2015bza, Dimastrogiovanni:2015pla, Kumar:2017ecc}, a particle $X$ can mediate primordial non-Gaussianities of observable size if it (1) has $m_X \sim \mathcal{O} (H_{\rm inf})$, (2) has sufficiently strong $X (\partial \phi) (\partial \phi)$ couplings, and (3) can give tree-level contribution to inflaton 3-point function which can come only from bosons. 

\subsubsection{Sinflaton}
\label{subsubsec:NG_sinflaton}

In the SUSY bi-axion model, mass of the sinflaton $(\eta_l)$ during inflation is
\begin{equation}
\label{eq:sinflaton_mass_sqrt6}
m_{\eta_l} \approx \sqrt{6} H_{\rm inf},
\end{equation}
which can be seen schematically from $V_{\rm SUGRA}$ as follows:
\begin{equation}
\label{eq:sinflaton_mass_from_curvature_coupling}
\begin{split}
V_{\rm SUGRA} &= e^K \left( \left| D_{\Phi_A} W \right|^2 + \left| D_{\Phi_B} W \right|^2 - 3 \left| W \right|^2 \right), \\
V(\eta_l) &\approx e^{\eta_l^2} V_{\rm inf} \ni \eta_l^2 V_{\rm inf} \approx 3 H_{\rm inf}^2 \eta_l^2 .
\end{split}
\end{equation}
This contribution to $m_{\eta_l}$ comes from the coupling of $\eta_l$ to the $\cancel{\textrm{SUSY}}$ curvature during inflation which also shows up in the KLR model as described in Section~\ref{sec:KL_toy_model} (see \eqref{eq:KL_model_sinflaton_mass}). However, as in the case of the KLR model (see \eqref{eq:KL_model_sinflaton_mass_additional}), $m_{\eta_l} \sim \mathcal{O}(H_{\rm inf})$ is not guaranteed in our SUSY bi-axion model too. A higher order term in K$\ddot{\mathrm{a}}$hler potential of the form
\begin{equation}
\label{eq:biaxion_sugra_cross_quartic_term}
K_5 \ni \delta (x_5) \frac{c_2}{\Lambda_2^2}(\Phi_A + \bar{\Phi}_A)^2 (\Phi_B + \bar{\Phi}_B)^2
\end{equation}
can give a contribution to the sinflaton mass as
\begin{equation}
\label{eq:biaxion_sugra_cross_quartic_term_sinflaton_mass_contri}
m_{\eta_l}^2 \approx \frac{2 V_{\rm inf}}{M_{\rm Pl}^2} + \frac{c_2 V_{\rm inf}}{\Lambda_2^2} = 6 H_{\rm inf}^2 \left( 1 + \frac{c_2 M_{\rm Pl}^2}{2 \Lambda_2^2} \right).
\end{equation}
Thus, for $\frac{c_2}{\Lambda_2^2} \gg \frac{1}{M_{\rm Pl}^2}$, $m_{\eta_l} \gg H_{\rm inf}$ is possible.\footnote{One might worry that the sub-leading term \eqref{eq:biaxion_sugra_cross_quartic_term} can have a dominant effect on the sinflaton mass and whether this signals breakdown of the EFT expansion. This is however not true. The sinflaton mass in \eqref{eq:sinflaton_mass_sqrt6} comes purely from $M_{\rm Pl}$-suppressed SUGRA contributions whereas the higher order K$\ddot{\mathrm{a}}$hler term \eqref{eq:biaxion_sugra_cross_quartic_term} gives a direct coupling with suppression scale ($\Lambda_2$) which can be below $M_{\rm Pl}$.} 
The effective higher order coupling \eqref{eq:biaxion_sugra_cross_quartic_term} between $\Phi_A$ and $\Phi_B$ can arise radiatively via the loops of hypermultiplet $(H_2, H_2^c)$ which is charged under both the gauge groups as $(1, N)$. Naive dimensional analysis suggests that this loop contribution to \eqref{eq:biaxion_sugra_cross_quartic_term} is
\begin{equation}
\label{eq:biaxion_sugra_cross_quartic_term_loop_contri}
\left(\frac{c_2}{\Lambda_2^2}\right)_{\rm loop} \sim \frac{g^2 N}{16 \pi^2} \frac{1}{m_{KK}^2}.
\end{equation}
Considering $N \sim \mathcal{O}(100)$ and $m_{KK} \sim M_5 \sim 0.1 M_{\rm Pl} $ \footnote{$M_5 \sim M_{\rm Pl} \left(\frac{1}{M_{\rm Pl} L}\right)^{1/3}$ is the scale at which gravity becomes strong in 5D.}, we can have $\left(\frac{c_2}{\Lambda_2^2}\right)_{\rm loop} \lesssim \frac{1}{M_{\rm Pl}^2}$ with $g \lesssim 0.1$. Thus, with $g \lesssim 0.1$, the contribution from loop-induced term \eqref{eq:biaxion_sugra_cross_quartic_term} to sinflaton mass is small, thus keeping $m_{\eta_l} \sim \mathcal{O} (H_{\rm inf})$, which is crucial to get observable NG mediated by it.

The references \cite{Kahn:2015mla, Ferrara:2015tyn} and \cite{Delacretaz:2016nhw} construct SUSY EFT of inflation with a minimal field content which does not include any scalar other than the inflaton, especially the sinflaton. This can be interpreted by the UV-completion of these EFTs having sinflaton with mass much greater than $H_{\rm inf}$. In our SUSY bi-axion model, as can be seen from \eqref{eq:biaxion_sugra_cross_quartic_term_sinflaton_mass_contri}, there exists a region of parameter space where $m_{\eta_l} \gg H_{\rm inf}$, which is consistent with the results of \cite{Kahn:2015mla, Ferrara:2015tyn, Delacretaz:2016nhw}. This parameter space corresponds to $\Lambda_2 \ll M_{\rm Pl}$ in \eqref{eq:biaxion_sugra_cross_quartic_term_sinflaton_mass_contri} or $g > 0.1$ in \eqref{eq:biaxion_sugra_cross_quartic_term_loop_contri}. This feature of having a region of parameter space allowing $m_{\rm sinflaton} \gg H_{\rm inf}$ is also present in the KLR model described in Section~\ref{sec:KL_toy_model} (see \eqref{eq:KL_model_sinflaton_mass_additional}).
However, in this case, the size of primordial NG suffers a severe exponential ``Boltzmann-suppression'' ($\sim e^{-\pi m_{\eta_l}/H_{\rm inf}}$). 
Below, we focus on the region where $m_{\eta_l} \sim \mathcal{O}(H_{\rm inf})$ which allows the sinflaton to be observable via primordial NG.

Even in the presence of $m_{\eta_l} \sim \mathcal{O} (H_{\rm inf})$, $\eta_l$ still needs to have sufficiently strong coupling with the inflaton to mediate NG of observable size. The $V_{\rm SUGRA}$ from \eqref{eq:biaxion_setup} has the following coupling which violates the shift symmetry for $\phi_l$ and hence is very small:
\begin{equation}
V_{\rm SUGRA} \ni 10^{-3} \frac{H_{\rm inf}}{M_{\rm Pl}} \ H_{\rm inf} \ \eta_l \phi_l^2.
\end{equation}
This coupling gives rise to the primordial NG of the following typical size \cite{Arkani-Hamed:2015bza, Kumar:2017ecc}:
\begin{equation}
\label{eq:fNL_sinflaton_within_VSUGRA}
f_{\rm NL} \sim 10^{-2} \frac{H_{\rm inf}}{M_{\rm Pl}} \lesssim 10^{-6}.
\end{equation}
This is much less than the sensitivity of the proposed experiments involving 21-cm cosmology, $f_{\rm NL} \sim 10^{-2}$ \cite{Loeb:2003ya, Munoz:2015eqa}, or even from more futuristic surveys, $f_{\rm NL} \sim 10^{-4}$ \cite{Meerburg:2016zdz}.

However, the following shift symmetry preserving, higher order, boundary-localized term in the K$\ddot{\mathrm{a}}$hler potential,
\begin{equation}
\label{eq:Phi_quartic_Kahler_term}
K_5 \ni \delta (x_5) \frac{c_1}{\Lambda_1^2} (\Phi_A + \bar{\Phi}_A)^4,
\end{equation}
can generate the following derivative coupling of sinflaton with inflaton:
\begin{equation}
\mathcal{L}_4 \ni \frac{c_1}{\Lambda_1^2} \eta_l^2 \left( \partial \phi_l \right)^2.
\end{equation}
The above coupling can give primordial NG of the size \cite{Arkani-Hamed:2015bza, Kumar:2017ecc}
\begin{equation}
\label{eq:fNL_sinflaton_from_additional_Kahler}
f_{\rm NL} \approx 0.03 \ c_1^2 \epsilon \left(\frac{M_{\rm Pl}}{\Lambda_1}\right)^4 \left(\frac{\left\langle \eta_l \right\rangle_{\rm inf}}{M_{\rm Pl}}\right)^2 \lesssim 10^{-6} \left(\frac{M_{\rm Pl}}{\Lambda_1}\right)^4,
\end{equation}
where the VEV of sinflaton during inflation is $\left\langle \eta_l \right\rangle_{\rm inf} \approx \frac{M_{\rm Pl}^2}{N f} \approx 0.1 M_{\rm Pl}$, as can be calculated from $V_{\rm SUGRA}$ using \eqref{eq:biaxion_setup}. $\epsilon$ in the above expression is the slow roll parameter of inflation which is constrained to be $\lesssim 10^{-2}$ \cite{Akrami:2018odb}.
The suppression scale $\Lambda_1$ in \eqref{eq:Phi_quartic_Kahler_term}, which would be the cutoff scale on the boundaries, has to be less than $M_5$. Considering $M_5 \sim \mathcal{O}(0.1) M_{\rm Pl}$, this implies that even for $\Lambda_1$ being very close to $M_5$, we can get $f_{\rm NL} \sim \mathcal{O}(10^{-2})$. This signal can be observed at the proposed 21-cm experiments as described after \eqref{eq:fNL_sinflaton_within_VSUGRA}. Furthermore, $\Lambda_1$ can be as low as the inflationary energy scale $V_{\rm inf}^{1/4} \lesssim 10^{-2} M_{\rm Pl}$, while maintaining EFT control, in which case $f_{\rm NL} \sim \mathcal{O}(1)$ or even higher is also possible.

\subsubsection{Boundary-localized gauge singlets}
\label{subsubsec:NG_brane_singlet}

It is also possible to see boundary-localized fields via primordial NG. Consider, for example, a chiral superfield $X$ localized at one of the boundaries and singlet under both the gauge groups $A$ and $B$. If it has the following K$\ddot{\mathrm{a}}$hler potential, i.e. a direct coupling with $\Phi_A$ preserving its shift symmetry,
\begin{equation}
\label{eq:boundary_localized_gauge_singlet_K}
K_5 \ni \delta (x_5) \left[ \frac{c_X}{\Lambda_X} (\Phi_A + \bar{\Phi}_A)^2 (X + \bar{X}) + \bar{X} X \right],
\end{equation}
then it has the following derivative interaction between the real scalar part of $X$ ($\eta_X$) and the inflaton:
\begin{equation}
\label{eq:boundary_localized_gauge_singlet_L}
\mathcal{L}_4 \ni \frac{c_X}{\Lambda_X} \eta_X \left( \partial \phi_l \right)^2.
\end{equation}
Also, analogous to the case of sinflaton as in \eqref{eq:sinflaton_mass_from_curvature_coupling}, mass of this gauge singlet during inflation is
\begin{equation}
\label{eq:boundary_localized_gauge_singlet_mass}
m_X \approx \sqrt{3} H_{\rm inf}.
\end{equation} 
\eqref{eq:boundary_localized_gauge_singlet_L} and \eqref{eq:boundary_localized_gauge_singlet_mass} imply that the size of primordial NG mediated by $\eta_X$ is as follows \cite{Arkani-Hamed:2015bza, Kumar:2017ecc}:
\begin{equation}
f_{\rm NL} \approx 0.75 \ c_X^2 \epsilon \left(\frac{M_{\rm Pl}}{\Lambda_X}\right)^2 \lesssim 10^{-2} \left(\frac{M_{\rm Pl}}{\Lambda_X}\right)^2.
\end{equation}
Similar to the case of $\Lambda_1$ as discussed below \eqref{eq:fNL_sinflaton_from_additional_Kahler}, $\Lambda_X \lesssim M_5 \sim 0.1 M_{\rm Pl}$ which can give $f_{\rm NL} \gtrsim \mathcal{O}(1)$. 

If the direct coupling in \eqref{eq:boundary_localized_gauge_singlet_K} is of the form $K_5 \ni \delta (x_5) \frac{c^\prime_X}{\Lambda_X^2} (\Phi + \bar{\Phi})^2 \bar{X} X$, then it gives the interaction $\mathcal{L}_4 \ni \frac{c^\prime_X}{\Lambda_X^2} |X|^2 \left( \partial \phi_l \right)^2$. In this case, the $f_{\rm NL}$ mediated by $X$ has an additional suppression factor due to its VEV during inflation: $f_{\rm NL} \sim c_X^2 \epsilon \left(\frac{M_{\rm Pl}}{\Lambda_X}\right)^2 \left(\frac{\langle X \rangle}{\Lambda_X}\right)^2 $. Hence, in order to get $f_{\rm NL}$ of an observable size, we need to have $\langle X \rangle$ during inflation to be sufficiently large as compared to $\Lambda_X$.

\subsection{Periodic modulations in the CMB}
\label{subsec:CMB_modulations}

Extra-dimensional realization of Natural Inflation gives the leading slowly varying inflaton potential with super-Planckian field range $(\sim f_{\rm eff} = Nf > M_{\rm Pl})$, while also generically giving sub-leading ``higher harmonic'' terms oscillating over a much shorter range $(\sim f, f/N \ll M_{\rm Pl})$. Although these higher harmonics in $V(\phi_{\rm inf})$ are suppressed by factors of $e^{-ML}$, they can still give observable effects in the form of primordial features with periodic modulations in the CMB power spectrum. These features, being motivated from various theoretical constructions, have been searched for in the Planck CMB data \cite{Wang:2002hf, Pahud:2008ae, Flauger:2009ab, Kobayashi:2010pz, Easther:2013kla, Flauger:2014ana, delaFuente:2014aca, Higaki:2014mwa, Choi:2015aem, Price:2015xwa}.

In our SUSY bi-axion model, there exist such higher harmonics in $V(\phi_l)$ arising from within the model, but they are small. These can come from the sub-dominant terms in the superpotential \eqref{eq:biaxion_SUGRA_superpotential_withBoundaryVEVchoice}, $\delta W  = \delta W \left(e^{-mL} \cdot e^{\frac{gL}{\sqrt{2} N} \Phi_l} \right)$, suppressed by powers of $e^{-mL}$. This  gives corrections to the inflaton potential of the form
\begin{equation}
\frac{\delta V}{V_{\rm inf}} \sim e^{-2nmL} \cos\left( n \frac{\phi_l}{Nf} \right).
\end{equation}
However, contributions to the periodic modulations in the CMB come only from the harmonics with $n \gg 1$ i.e. $n \sim \mathcal{O}(N)$. But, such $\cos\left(\mathcal{O}(N) \cdot \frac{\phi_l}{Nf}\right)$ terms in the potential are hugely suppressed by $\sim e^{-2mL \cdot \mathcal{O}(N)}$. Thus, the higher harmonics from within the SUSY bi-axion model cannot give rise to observable CMB periodic modulations.

But, let us now consider contribution from a generic heavy hypermultiplet beyond our minimal model $(\mathcal{H}_3, \mathcal{H}_3^c)$, with mass $M$ and charges $(n_A, n_B)$ under the gauge groups $A$ and $B$. This will give an additional term in the superpotential \eqref{eq:biaxion_SUGRA_superpotential_withBoundaryVEVchoice} as
\begin{equation}
\label{eq:CMB_modulations_heavy_charge_W}
\delta W \approx 2 v^2 e^{-ML} \left[ 1 - e^{\frac{gL}{\sqrt{2}} (n_A \Phi_A + n_B \Phi_B)} \right],
\end{equation}
where we have taken the parameters governing boundary VEVs of $H_3, H_3^c$ to be equal: $v_3 = v_3^\prime = v$. As expected, this is suppressed by $e^{-ML}$ which is ``filtered'' out by the extra dimension for $M \gg 1/L$. However, as discussed below, the precision CMB observables can be sensitive to the contributions to periodic modulations sourced by such a hypermultiplet if it is not too heavy. 
The contribution to the inflaton potential from \eqref{eq:CMB_modulations_heavy_charge_W} is
\begin{equation}
\frac{\delta V}{V_{\rm inf}} \approx n_B e^{2mL} e^{-ML} \cos\left[(N n_B - n_A) \frac{\phi_l}{Nf}\right].
\end{equation}
The observational constraint on the size of CMB periodic modulations is $\left|\frac{\delta V}{V_{\rm inf}} \right| \lesssim 10^{-5}$, also depending upon the higher harmonic frequency \cite{Choi:2015aem}. Considering $n_B \sim \mathcal{O}(N) \sim 100$ and $e^{-mL} \sim 1/3$, we can get $\left|\frac{\delta V}{V_{\rm inf}} \right| \sim 10^{-5}$ from $M \sim 20 \times \frac{1}{L}$. This shows sensitivity of CMB periodic modulations to the charged matter much heavier than the KK scale!

The 5D gauge theory being non-renormalizable has a cutoff which is given by 
$\Lambda_{\rm 5D} \sim \frac{c}{g^2} \frac{1}{L}$. 
As discussed below \eqref{eq:biaxion_sugra_cross_quartic_term_loop_contri}, we require $g \lesssim 0.1$ in order to have $m_{\eta_l} \sim \mathcal{O}(H_{\rm inf})$ for getting observable primordial NG mediated by sinflaton. Hence, for $g \lesssim 0.1$ and $c \sim \mathcal{O}(1)$, the cutoff is $\Lambda_{\rm 5D} \gtrsim 100 \times \frac{1}{L}$. 
Thus, charged matter beyond the minimal model with 
\begin{equation}
M \lesssim \frac{1}{5} \Lambda_{\rm 5D}
\end{equation}
can generate observable periodic modulations in the CMB power spectrum with $\left|\frac{\delta V}{V_{\rm inf}} \right| \sim 10^{-5}$. Of course, some such heavy states are expected near the cutoff of 5D gauge theory as part of a UV-completion of our non-renormalizable effective field theory.

\section{Conclusions}
\label{sec:conclusions}

In the present work, we demonstrated the compatibility of low-energy SUSY (i.e. SUSY broken only at somewhat above the EW scale) with high-scale axionic inflation where the axionic nature of inflaton is derived from extra-dimensional gauge symmetry. The  inflaton potential, in the presence of SUSY, can be generated at tree-level by charged matter in the 5D bulk with gauge symmetry breaking at the 5D boundaries. 
We also required that this robust gauge-theoretic origin for the inflaton satisfy the Weak Gravity Conjecture quantum gravity constraints, which are especially tight given the  super-Planckian inflaton field range required by the data (Lyth bound). But we showed that this can be achieved by introducing two axion supermultiplets, containing a light inflaton direction having $f_{\rm eff} > M_{\rm Pl}$. The heavy sector, apart from stabilizing the inflationary trajectory, also contributes dominantly to SUSY breaking ($\cancel{\textrm{SUSY}}$) during inflation. The Goldstino of spontaneous $\cancel{\textrm{SUSY}}$ during inflation lies mostly in this heavy sector. 

Our SUSY bi-axion model displays an interesting interplay of electroweak (EW), cosmological constant (CC) and superpotential ($W_0$) fine-tunings after considering the $\cancel{\textrm{SUSY}}$ vacuum we occupy today. The fine-tuning for EW and CC, as usual, prefer low-scale $\cancel{\textrm{SUSY}}$. The $W_0$ fine-tuning, however, shows preference for high-scale $\cancel{\textrm{SUSY}}$. We showed that the net fine-tuning is dominated by EW and CC fine-tunings and hence prefers low-scale $\cancel{\textrm{SUSY}}$ i.e. somewhat above the EW scale.

The observable signals in our model can come in the form of primordial non-Gaussianities (NG) and periodic modulations in the CMB. The sinflaton can naturally have $\mathcal{O} (H_{\rm inf})$ mass via its coupling to the $\cancel{\textrm{SUSY}}$ curvature during inflation. It can also naturally have sufficiently strong couplings with inflaton such that it can be seen via primordial NG in  future 21-cm experiments, with the measure of NG, $f_{\rm NL}$, being $\gtrsim 10^{-2}$. The sinflaton mass can receive large contributions from higher order K$\ddot{\mathrm{a}}$hler terms which, however, can be kept sub-dominant with small enough gauge coupling. 
Similarly, a boundary-localized gauge singlet can have $\mathcal{O} (H_{\rm inf})$ mass during inflation and strong enough coupling with inflaton, via higher order K$\ddot{\mathrm{a}}$hler couplings, thus allowing it to mediate large primordial NG with even $f_{\rm NL} \gtrsim \mathcal{O}(1)$.

Although the extra dimension acts as a ``filter'' for the unknown UV-completion of our non-renormalizable model, with $e^{-ML}$ suppression, the precision observables in the CMB can still probe modulating features imprinted by such heavy physics. We showed that charged matter, not far below the effective field theory cutoff of our model, can generate modulations in the inflationary potential, $\left|\frac{\delta V}{V_{\rm inf}} \right| \sim 10^{-5}$, which lie within the sensitivity of ongoing searches \cite{Flauger:2014ana, Choi:2015aem}.

As mentioned in Section~\ref{sec:intro}, the recent Planck 2018 CMB data \cite{Akrami:2018odb} puts tight constraints on Natural Inflation. The 
bi-axionic inflation studied here, while very roughly giving a Natural Inflation potential, can have significant differences at precision level that can be used to better agree with the data, as exemplified in \cite{Peloso:2015dsa}. We hope to further explore SUSY axionic inflation models in the future for the best fit to the precision data. 

\appendix
\section{SUGRA preliminaries}
\label{sec:appendix_sugra_basics}

We write here the important SUGRA expressions relevant for the present paper. See \cite{Wess:1992cp} for review and further details.

For a general K$\ddot{\mathrm{a}}$hler potential and superpotential for chiral superfields $\Phi_i$,
\begin{equation}
K = K(\Phi_i, \bar{\Phi}_i) \ , \ W = W(\Phi_i),
\end{equation}
the SUGRA scalar potential is
\begin{equation}
\label{eq:appendix_V_SUGRA}
V_{\textrm{scalar}}(\Phi_i, \bar{\Phi}_i) = e^K \left[ K^{-1}_{\Phi_i \bar{\Phi}_j} \left(D_{\Phi_i} W \right) \left(D_{\bar{\Phi}_j} \overline{W}\right) - 3 W \overline{W} \right],
\end{equation}
with subscripts referring to the respective partial derivatives, and with
\begin{equation}
D_{\Phi_i} W \equiv W_{\Phi_i} + K_{\Phi_i} W.
\end{equation}
$\langle D_{\Phi_i} W \rangle$ is the SUSY order parameter for each of the superfields $\Phi_i$. If there exists spontaneous breaking of SUSY in a model, it gives rise to a massless Goldstino,  
\begin{equation}
\psi_{\textrm{Goldstino}} \propto \langle D_{\Phi_i} W \rangle \ \psi_{\Phi_i},
\end{equation}
where $\psi_{\Phi_i}$ are fermions in the superfields $\Phi_i$. The Goldstino is further ``eaten'' by the gravitino which then becomes massive. This is called the ``super-Higgs mechanism''.

\acknowledgments

The authors would like to thank Soubhik Kumar, Marco Peloso and Jesse Thaler for helpful discussions. This research was supported in part by the NSF under Grant No. PHY-1620074 and by the Maryland Center for Fundamental Physics (MCFP).


\bibliographystyle{JHEP}
\bibliography{SUSY_inflation_from_5D}

\end{document}